\documentclass[titlepage,twoside,10pt,letter]{article}

\usepackage{times}
\usepackage{amsmath}
\usepackage{amsfonts}
\usepackage{amssymb}
\usepackage{graphicx}
\usepackage{mathrsfs}
\usepackage{tabularx}
\usepackage{caption}
\usepackage{bbold}
\usepackage{color}
\usepackage{fancybox}
\usepackage{verbatim}
\usepackage{hyperref}
\usepackage{tikz}
\usepackage{float}
\usetikzlibrary{arrows,automata}
\usetikzlibrary{trees}

\def\bR{\mathbb{R}}

\def\b1{\mathbb{1}}

\def\eE{\mathsf{E}}

\def\tT{\mathtt{T}}

\def\cH{{\mathcal{H}}}

\def\cL{\mathcal{L}}

\def\e1{\mathsf{1}}

\def\ofx{(x)}

\def\of0{(0)}

\def\d{\partial}

\def\bf0{\mathbf{0}}
\def\cp1{\mathbb{CP}^1}

\def\arg{\mathrm{arg}}

\begin{document}

\title{\textbf{Epidemic control\\
via stochastic optimal control}}
\author{\textbf{Andrew Lesniewski}\\
Department of Mathematics\\
Baruch College\\
55 Lexington Avenue\\
New York, NY 10010
\date{Draft of \today\\
First draft: April 14, 2020}}
\maketitle

\begin{abstract}
We study the problem of optimal control of the stochastic SIR model. Models of this type are used in mathematical epidemiology to capture the time evolution of highly infectious diseases such as COVID-19. Our approach relies on reformulating the Hamilton-Jacobi-Bellman equation as a stochastic minimum principle. This results in a system of forward backward stochastic differential equations, which is amenable to numerical solution via Monte Carlo simulations. We present a number of numerical solutions of the system under a variety of scenarios.
\end{abstract}

\section{\label{intro_sec}Introduction}

The classic SIR model, originally proposed in \cite{KM27}, is the standard tool of mathematical epidemiology \cite{BCF19} for quantitative analysis of the spread of an epidemic. It describes the state of the affected population in terms of three state variables, traditionally denoted by $S,I$, and $R$:
\begin{itemize}
\item[(i)]{$0\leq S\leq 1$, the fraction of individuals who are \textit{susceptible} to the disease.}
\item[(ii)]{$0\leq I\leq 1$, the fraction of individuals who are \textit{infected} with the disease.}
\item[(iii)]{$0\leq R\leq 1$, the fraction of individuals who have been \textit{removed} and are immune to the disease. Note that, in this simplified model, $R$ includes the individuals who have left the population through mortality.}
\end{itemize}

The model dynamics is given as the following dynamical system:
\begin{equation}\label{sirdyn}
\begin{split}
\frac{\dot S(t)}{S(t)}&=-\beta I(t),\\
\frac{\dot I(t)}{I(t)}&=\beta S(t)-\gamma,\\
\frac{\dot R(t)}{I(t)}&=\gamma,
\end{split}
\end{equation}
where the constant parameters $\beta>0$ and $\gamma>0$ are called the \textit{infection rate} and \textit{recovery rate}, respectively. The evolution above is subject to the initial conditions:
\begin{equation}\label{initcond}
\begin{split}
S(0)&=S_0,\\
I(0)&=I_0,\\
R(0)&=1-S_0-I_0.
\end{split}
\end{equation}
Notice that this dynamics obeys the conservation law
\begin{equation}\label{sumofvarbs}
S(t)+I(t)+R(t)=1,
\end{equation}
consistent with the assumption that the variables $S,I$, and $R$ represent  population fractions. This means that the variable $R$ is, in a way, redundant, as its current value does not affect the dynamics of $S$ and $I$, and it can be computed in a straightforward manner from \eqref{sumofvarbs}.
It is thus natural to consider the two dimensional system defined by $S,I$ only.

Eventually every epidemic comes to a natural halt\footnote{In the SIR model \eqref{sirdyn} this is reflected by the fact that the solution $(S(t),I(t))=(0,0)$ is globally asymptotically stable.}, but its impact on the population may be very serious. The overall objective of epidemic control is to slow down the spread of infection in a way that it does not overwhelm the healthcare system and it allows the economy to function. All of this should be done within the limits of available resources.

In this note we study the problem of optimal control of an epidemic modeled by means of a stochastic extension of the SIR model (see Section \ref{ssir_sec} for definition). We assume that the controlling agent (``government'') has the ability to impact the spread of the disease through one of the policies (or the combination of both):
\begin{itemize}
\item[(i)]{Vaccination of susceptible individuals, with rate $v$. This makes the fraction $vS$ of the susceptible population immune to the disease.}
\item[(ii)]{Isolation of infected individuals, with rate $i$. This removes the fraction $iI$ of the infected population and prevents it from spreading the disease.}
\end{itemize}
The controlled dynamics of the SIR model reads \cite{AM79}, \cite{HD11}:
\begin{equation}\label{consirdyn}
\begin{split}
\frac{\dot S(t)}{S(t)}&=-\beta I(t)-v(t),\\
\frac{\dot I(t)}{I(t)}&=\beta S(t)-\gamma-i(t),\\
\dot R(t)&=v (t)S(t)+(\gamma+i(t))I(t).
\end{split}
\end{equation}

For efficiency, we will be using the notation $X_1=S$ and $X_2=I$ throughout the remainder of the paper.

Mathematical models are only as good as (i) their analytic specifications, and (ii) the data that fuel them. During initial phases of an epidemic the available data tend to be of limited usefulness: because of the lack of reliable large scale testing, it is not really known what fractions of the population fall into the different compartments $S$, $I$, and $R$. This may lead to a panic reaction of the population and a chaotic and economically devastating public health response to the epidemic. In the absence of an effective vaccine (which would allow to immunize a portion of the population) the optimal policy is to isolate at least a significant fraction of the infected individuals so that the basic reproduction ratio $R_0$ can be brought significantly below one. Since we lack the knowledge who is infected and who is not, the public health response is to try to isolate everyone, whether susceptible, infected or immune.

These circumstances impose a serious limitation on practical applicability of the approach to optimal epidemic control discussed in this paper, as well as other quantitative approaches. Unless the inputs to the model ($\beta$, $\gamma$, and the current snapshots of $S$ and $I$) can reliably be estimated from the available data, the model's output is unreliable\footnote{``This is indeed a mystery,'' Watson remarked. ``What do you imagine that it means?''\\	
``I have no data yet,'' Holmes replied. ``It is a capital mistake to theorise before one has data. Insensibly one begins to twist facts to suit theories, instead of theories to suit facts.'' \cite{ACD}}.

The paper is organized as follows. In Section \ref{ssir_sec} we review the formulation of the continuous time stochastic SIR model. The optimal control problem for the stochastic SIR model is formulated in Section \ref{hjb_sec}. The optimal control problem is recast as the stochastic minimum principle problem and formulated in terms of a system of forward backward stochastic differential equations (FBSDE). We present an algorithm for solving this system in Section \ref{numer_sec}. This section presents also the results of a number of numerical experiments involving the three mitigation policies within various cost function regimes.

\noindent
\textbf{Acknowledgement.} I would like to thank Nicholas Lesniewski for numerous discussions.

\section{\label{ssir_sec}Stochastic SIR model}

We consider a continuous time stochastic extension of the deterministic SIR model, see \cite{BBSG17} and references therein. Let $W_t$ denote the standard Brownian motion, and let $\dot W_t$ denote the white noise process. We assume that the infection rate $\beta$ is subject to random shocks, rather than being constant, namely
\begin{equation}\label{sbeta}
\beta_t=\beta+\sigma\dot W_t,
\end{equation}
while the recovery rate $\gamma$ remains constant. Here $\sigma>0$ is a constant volatility parameter. This leads to the following system of stochastic differential equations (SDE), driven by $W_t$
\begin{equation}\label{onefactmod}
\begin{split}
\frac{d X_{1,t}}{ X_{1,t}}&=-\beta  X_{2,t} dt-\sigma  X_{2,t} dW_t,\\
\frac{d X_{2,t}}{ X_{2,t}}&=(\beta  X_{1,t}-\gamma)dt+\sigma X_{1,t} dW_t,
\end{split}
\end{equation}
with the initial conditions
\begin{equation}\label{sinitcond}
\begin{split}
X_{1,0}&=S_0,\\
X_{2,0}&=I_0.
\end{split}
\end{equation}

The third component of the process, $X_3=R$, follows the dynamics
\begin{equation}
\frac{d X_{3,t}}{ X_{2,t}}=\gamma dt,
\end{equation}
which implies that the conservation law
\begin{equation}\label{stochconslaw}
X_t+Y_t+Z_t=1
\end{equation}
continues to hold in the stochastic model.

Notice that, under the stochastic SIR model \eqref{onefactmod}, an epidemic eventually comes to a natural end. More precisely, the solution $(X_{1,t},X_{2,t})=(0,0)$ to \eqref{onefactmod} is stable in probability \cite{K12}. In order to see it, we set
\begin{equation}
V_\rho(X_1,X_2)=\beta(X_1+X_2)-\gamma\log(1+\rho X_1),
\end{equation}
for $0\leq X_i\leq 1$, $i=1,2$, and fixed $0<\rho<\beta/\gamma$. Then $V_\rho(0,0)=0$, and $V_\rho(X_1,X_2)>0$ in a neighborhood of $(0.0)$. Furthermore, denoting by $\cL$ the generator of the process\eqref{onefactmod}, we verify that
\begin{equation}
\begin{split}
\cL V_\rho(X_1,X_2)&=-\beta\gamma\,\frac{I}{\rho X_1+1}-\frac{\gamma\sigma\rho^2}{2}\,\frac{X_1 X_2}{(1+\rho X_1)^2}\\
&\leq 0.
\end{split}
\end{equation}
In other words, $V_\rho(X_1,X_2)$ is a Lyapunov function for \eqref{onefactmod} and our claim follows from Theorem 5.3 in \cite{K12}.

The model \eqref {onefactmod} is a one factor model, driven by a single source of randomness. There is a natural two-factor version of the stochastic SIR model \cite{LL20}, in which also the recovery rate $\gamma$ is allowed to be subject to white noise shocks. For simplicity, our analysis will focus on the one-factor model \eqref{onefactmod}.

\section{\label{hjb_sec}HJB equation}

We frame the problem of epidemic control as a stochastic control problem \cite{B57}. We denote by $u=(u_1,u_2)\equiv(v,i)$ the vaccination and isolation controls, and we denote the controlled process by $X^u_t$. Generalizing the deterministic specification \eqref{consirdyn} to the stochastic case, we assume that the dynamics of $X^u_t$ is given by:
\begin{equation}\label{contr_sir1f}
\begin{split}
\frac{d X^u_{1,t}}{X^u_{1,t}}&=-(\beta X^u_{2,t}+u_{1,t}) dt-\sigma X^u_{2,t} dW_t,\\
\frac{d X^u_{2,t}}{X^u_{2,t}}&=(\beta X^u_{1,t}-\gamma-u_{2,t})dt+\sigma X^u_{1,t} dW_t,
\end{split}
\end{equation}
subject to initial conditions \eqref{sinitcond}.

Two special cases of the controlled process are of interest. If a vaccine against the disease is unavailable, we set $u_1=0$ in the equation above, which yields the following controlled process:
\begin{equation}\label{contr_sir1fi}
\begin{split}
\frac{d X^u_{1,t}}{X^u_{1,t}}&=-\beta X^u_{2,t} dt-\sigma X^u_{2,t} dW_t,\\
\frac{d X^u_{2,t}}{X^u_{2,t}}&=(\beta X^u_{1,t}-\gamma-u_{2,t})dt+\sigma X^u_{1,t} dW_t.
\end{split}
\end{equation}
We will refer to this policy as an isolation policy.

Similarly, we can consider a vaccination policy, for which $u_2=0$. In this case the controlled dynamics reads
\begin{equation}\label{contr_sir1f}
\begin{split}
\frac{d X^u_{1,t}}{X^u_{1,t}}&=-(\beta X^u_{2,t}+u_{1,t}) dt-\sigma X^u_{2,t} dW_t,\\
\frac{d X^u_{2,t}}{X^u_{2,t}}&=(\beta X^u_{1,t}-\gamma)dt+\sigma X^u_{1,t} dW_t.
\end{split}
\end{equation}

We assume a finite time horizon $T<\infty$. The controlling agent's objective is to minimize a running cost function $c(X_t,u_t)$ and the terminal value function $G(X_T)$. In other words, we are seeking a policy $u^\ast$ such that
\begin{equation}\label{optContr}
u^\ast=\mathop{\arg\min}_u\eE\Big(\int_0^T c(X_t^u,u_t) dt+G(X_T)\Big).
\end{equation}

We consider the following cost function:
\begin{equation}
c(x, u)=c_1(x_1,u_1)+c_2(x_2,u_2),
\end{equation}
where
\begin{equation}
c_i(x_i,u_i)=\big(\frac12\,L_i u_i^2+M_i u_i + N_i\big) x_i,
\end{equation}
for $i=1,2$. In other words, the running cost of vaccination is assumed to be proportional to the number of susceptible individuals, while the cost of isolation is assumed to be proportional to the number of infected individuals. The coefficients $L_i>0, M_i, N_i$ are determined by the overall cost of following the mitigation policy. In particular, they should be selected so that the running cost functions are strictly positive.

As the terminal value function we take the transmission rate of the infection \cite{HD11}, namely
\begin{equation}
G(x)=\beta x_1 x_2.
\end{equation}

We now invoke stochastic dynamic programming, see eg. \cite{FS92}, \cite{P09}. The key element of this approach is the value function $J(t,x,y)$. It is determined by two requirements:
\begin{itemize}
\item[(B1)]{It satisfies Bellman's principle of optimality,
\begin{equation}
J(t,X_t)=\min_{u}\;\eE_t\big(J(t+dt,X_{t+dt})+c(X_t,u_t)dt\big),
\end{equation}
for all $0\leq t<T$, where $\eE_t$ denotes conditional expectation with respect to the information set at time $t$, and where the minimum is taken over all admissible controls $u_t$, \cite{FS92}.}
\item[(B2)]{It satisfies the terminal condition,
\begin{equation}
J(T,X_T)=G(X_T).
\end{equation}}
\end{itemize}

Using Ito's lemma, we verify that these two conditions lead to the following nonlinear partial differential equation for the value function, namely the stochastic Hamilton-Jacobi-Bellman equation:
\begin{equation}\label{hjb11f}
\begin{split}
\dot{J}&+\min_u\,\Big\{-x_1(\beta x_2+u_1) J_{x_1}+(\beta x_1-\gamma-u_2){x_2} J_{x_2}\\
&+\frac12\,\sigma^2 x_1^2 x_2^2\big(J_{x_1 x_1}+J_{x_1 x_2}-2J_{x_1 x_2}\big)+c(x,u)\Big\}=0,
\end{split}
\end{equation}
subject to the terminal condition
\begin{equation}
J(T,x)=G(x).
\label{eq:defValueFunctionTermimal}
\end{equation}

As the first step towards solving the HJB equation \eqref{hjb11f}, we let $u=u^\ast$ denote the minimizer of the expression inside the curly parentheses in \eqref{hjb11f}. In other words, $u^\ast$ satisfies
\begin{equation*}
-x_i J_{x_i}+(L_i u_i+M_i)x_i=0,
\end{equation*}
which leads to the following first order condition on $u^\ast$:
\begin{equation}\label{uast1f}\
u^\ast_i=\frac{1}{L_i}\,(J_{x_i}-M_i).
\end{equation}
Substituting $u^*$ back to the HJB equation yields the equation
\begin{equation}\label{hjb21f}
\begin{split}
\dot{J}-\beta x_1 x_2 J_{x_1}&+(\beta x_1-\gamma)x_2 J_{x_2}+\frac12\sigma^2 x_1^2 x_2^2\big(J_{x_1 x_1}+J_{x_1 x_2}-2J_{x_1 x_2}\big)\\
&\big(N_1-\frac{(J_{x_1}-M_1)^2}{2L_1}\big)x_1+\big(N_2-\frac{(J_{x_2}-M_2)^2}{2L_2}\big)x_1=0.
\end{split}
\end{equation}

We do not believe that the solution to this equation can be explicitly represented in terms of standard functions. This is a three dimensional partial differential equation, and solving it numerically may pose challenges. Rather than following this path, we shall invoke the stochastic minimum principle and reformulate the problem as a system of FBSDEs. Among the advantages of this approach is that it might be amenable to a deep learning approach via the method advocated in \cite{HJE18}.

\section{\label{smp_sec}Stochastic minimum principle}

The stochastic minimum principle, see \cite{P09} and references therein, offers an alternative approach to stochastic optimal control. It is a stochastic version of Pontryagin's minimum principle introduced in the context of deterministic optimal control \cite{PBGM92}. It also offers an effective numerical methodology to solving the HJB equation \eqref{hjb21f} via Monte Carlo simulations. In this approach, the key object is the Hamiltonian function $\cH=\cH(x,u,y,z)$ of four arguments and a system of stochastic differential equations, both forward and backward, determined by $\cH$.

Specifically, for the case of the controlled stochastic SIR model \eqref{contr_sir1f}, we have $x,u,y,z\in\bR^2$, and the Hamiltonian function reads:
\begin{equation}
\begin{split}
\cH(x,u,y,z)=&-x_1(\beta x_2+u_1)y_1+(\beta x_1-\gamma-u_2)x_2 y_2\\
&+\sigma x_1 x_2(z_2-z_1)+c(x,u).
\end{split}
\end{equation}
We consider the following system of stochastic Hamilton's equations:
\begin{equation}\label{hamXEqs}
\begin{split}
dX_t&=\nabla_y \cH(X_t,u_t,Y_t,Z_t) dt+\sigma (X_t) dW_t,\\
X_0&=
\begin{pmatrix}
S_0\\
I_0
\end{pmatrix},
\end{split}
\end{equation}
and
\begin{equation}\label{hamYEqs}
\begin{split}
-dY_t&=\nabla_x \cH(X_t,u_t,Y_t,Z_t) dt-Z_t dW_t,\\
Y_T&=\nabla_x G(X_T),
\end{split}
\end{equation}
where
\begin{equation}
\sigma(X_t)=\sigma X_{1,t}X_{2,t}
\begin{pmatrix}
-1\\
1
\end{pmatrix}.
\end{equation}
Equation \eqref{hamXEqs} is merely an alternative way of writing the underlying controlled diffusion process \eqref{contr_sir1f}, while equation \eqref{hamYEqs} reflects the dynamics of the control variables given the running cost function. Note that while the first of the equations \eqref{hamXEqs} is a standard (forward) SDE, the second one is a backward stochastic differential equation (BSDE), see e.g. \cite{EPQ97}.

Explicitly, these four equations can be stated as
\begin{equation}\label{fbsde}
\begin{split}
dX_{1,t}&=-X_{1,t}(\beta X_{2,t}+u_{1,t}) dt-\sigma X_{1,t} X_{2,t} dW_t,\\
dX_{2,t}&=(\beta X_{1,t}-\gamma-u_{2,t}) X_{2,t} dt+\sigma  X_{1,t} X_{2,t} dW_t,\\
-dY_{1,t}&=\big(\beta(Y_{2,t}-Y_{1,t})+\sigma (Z_{2,t}-Z_{1,t}))X_{2,t}-u_{1,t}Y_{1,t}\\
&\qquad\qquad\qquad\qquad+c_{x_1}(X_t,u_t)\big) dt-Z_{1,t} dW_t,\\
-dY_{2,t}&=\big(\beta(Y_{2,t}-Y_{1,t})+\sigma(Z_{2,t}-Z_{1,t})\big)X_{1,t}-(\gamma+u_{2,t}) Y_{2,t}\\
&\qquad\qquad\qquad\qquad+c_{x_2}(X_t,u_t)\big) dt-Z_{2,t} dW_t,
\end{split}
\end{equation}
subject to the boundary conditions in \eqref{hamXEqs}.

Let now choose $u^\ast$ to be the policy such that $X_t^\ast\equiv X_t^{u^\ast}$ is a solution to \eqref{fbsde} and
\begin{equation}
\cH(X^\ast_t,u^\ast_t,Y^\ast_t,Z^\ast_t)=\min_{u}\cH(X_t^u,u_t,Y_t^u,Z_t^u),
\end{equation}
Then $u^\ast$ is an optimal control, i.e. it satisfies the optimality condition \eqref{optContr}.

In fact, there is a direct link between the Hamilton function $\cH$ and the value function $J$ (also known in classical dynamics as the action function). Namely, we set \cite{P09}, \cite{YZ99}:
\begin{equation}
J(t, X_t)=\int_0^t Y_s^\tT dX_s-\cH(X_s,u_s,Y_s,Z_s)ds.
\end{equation}
If $u^\ast$ is an optimal control and $X^\ast$ denotes the corresponding optimal diffusion process, then the pair
\begin{equation}
\begin{split}\label{yz_expl}
Y_t&=\nabla_x J(t, X_t^\ast),\\
Z_t&=\nabla_x^2 J(t, X_t^\ast)\,\sigma(X_t^\ast)
\end{split}
\end{equation}
is the solution to the BSDE \eqref{hamYEqs}.

Going back to the main line of reasoning, we find that $u^\ast$ has to satisfy
\begin{equation}
u^\ast_i=\frac{y_i-M_i}{L_i}\,.
\end{equation}
From equations \eqref{yz_expl} we see that, up to a simple linear transformation, $Y_t$ is essentially the optimal policy. Furthermore, the process $Z_t$ can be thought of as the sensitivity of the optimal policy to the underlying process $X_t$ (multiplied by the instantaneous volatility of that process).

Substituting this expression into \eqref{fbsde}, we find that the optimal process has to follow the dynamics:
\begin{equation}\label{fbsde_opt}
\begin{split}
dX_{1,t}^\ast&=-X_{1,t}^\ast\big(\beta X_{2,t}^\ast+\frac{1}{L_1}\,(Y_{1,t}^\ast-M_1)\big) dt-\sigma X_{1,t}^\ast X_{2,t}^\ast dW_t,\\
dX_{2,t}^\ast&=\big(\beta X_{1,t}^\ast-\gamma-\frac{1}{L_2}\,(Y_{2,t}^\ast-M_2)\big) X_{2,t}^\ast dt+\sigma  X_{1,t}^\ast X_{2,t}^\ast dW_t,\\
-dY_{1,t}^\ast&=\big(\beta(Y_{2,t}^\ast-Y_{1,t}^\ast)+\sigma (Z_{2,t}^\ast-Z_{1,t}^\ast))X_{2,t}^\ast dt\\
&\qquad\qquad\qquad+\big(N_1-\frac{1}{2 L_1}\,(Y_{1,t}^\ast-M_1)^2\big) dt-Z_{1,t}^\ast dW_t,\\
-dY_{2,t}^\ast&=\big(\beta(Y_{2,t}^\ast-Y_{1,t}^\ast)+\sigma(Z_{2,t}^\ast-Z_{1,t}^\ast)\big)X_{1,t}^\ast dt\\
&\qquad\qquad+\big(-\gamma Y_{2,t}^\ast+N_2-\frac{1}{2 L_2}\,(Y_{2,t}^\ast-M_2)^2\big) dt-Z_{2,t}^\ast dW_t,
\end{split}
\end{equation}
subject to the boundary conditions in \eqref{hamXEqs} and \eqref{hamYEqs}. In particular, the isolation only and vaccination only policies are given by the following systems of FBSDEs:
\begin{equation}\label{fbsde_opt_i}
\begin{split}
dX_{1,t}^\ast&=-\beta X_{1,t}^\ast X_{2,t}^\ast dt-\sigma X_{1,t}^\ast X_{2,t}^\ast dW_t,\\
dX_{2,t}^\ast&=\big(\beta X_{1,t}^\ast-\gamma-\frac{1}{L_2}\,(Y_{2,t}^\ast-M_2)\big) X_{2,t}^\ast dt+\sigma  X_{1,t}^\ast X_{2,t}^\ast dW_t,\\
-dY_{1,t}^\ast&=\big(\beta(Y_{2,t}^\ast-Y_{1,t}^\ast)+\sigma (Z_{2,t}^\ast-Z_{1,t}^\ast))X_{2,t}^\ast dt-Z_{1,t}^\ast dW_t,\\
-dY_{2,t}^\ast&=\big(\beta(Y_{2,t}^\ast-Y_{1,t}^\ast)+\sigma(Z_{2,t}^\ast-Z_{1,t}^\ast)\big)X_{1,t}^\ast dt\\
&\qquad\qquad+\big(-\gamma Y_{2,t}^\ast+N_2-\frac{1}{2 L_2}\,(Y_{2,t}^\ast-M_2)^2\big) dt-Z_{2,t}^\ast dW_t,
\end{split}
\end{equation}
and
\begin{equation}\label{fbsde_opt_v}
\begin{split}
dX_{1,t}^\ast&=-X_{1,t}^\ast\big(\beta X_{2,t}^\ast+\frac{1}{L_1}\,(Y_{1,t}^\ast-M_1)\big) dt-\sigma X_{1,t}^\ast X_{2,t}^\ast dW_t,\\
dX_{2,t}^\ast&=\big(\beta X_{1,t}^\ast-\gamma\big) X_{2,t}^\ast dt+\sigma  X_{1,t}^\ast X_{2,t}^\ast dW_t,\\
-dY_{1,t}^\ast&=\big(\beta(Y_{2,t}^\ast-Y_{1,t}^\ast)+\sigma (Z_{2,t}^\ast-Z_{1,t}^\ast))X_{2,t}^\ast dt\\
&\qquad\qquad\qquad+\big(N_1-\frac{1}{2 L_1}\,(Y_{1,t}^\ast-M_1)^2\big) dt-Z_{1,t}^\ast dW_t,\\
-dY_{2,t}^\ast&=\big(\beta(Y_{2,t}^\ast-Y_{1,t}^\ast)+\sigma(Z_{2,t}^\ast-Z_{1,t}^\ast)\big)X_{1,t}^\ast dt-\gamma Y_{2,t}^\ast dt-Z_{2,t}^\ast dW_t,
\end{split}
\end{equation}
respectively.

Even though derived in the context of a meaningful underlying dynamics, there is no \textit{a priori} reason why these equations should have solutions. The drivers of the backward equations above contain terms quadratic in $Y$, and so the standard existence theorems \cite{EPQ97} do not apply. We proceed in the following assuming that these systems do, in fact, have solutions. 

\section{\label{numer_sec}Numerical solution}

In this section we discuss a numerical algorithm for solving the system \eqref{fbsde_opt}. Applying this methodology in a number of numerical experiments, we present compelling evidence that solutions to \eqref{fbsde_opt} exist and are meaningful over a wide range of model parameters.

\subsection{\label{discFBSDE}Simulating the optimal FBSDE}

We start by describing a method for discretization of the basic FBSDE system \eqref{fbsde_opt}. We notice first that the two dimensional system defined by the state variables $S,I$ is in fact Hamiltonian \cite{LR04}. Namely, we define new (canonical) variables
\begin{equation}
\begin{split}
q&=-\log S,\\
p&=-\log I,
\end{split}
\end{equation}
and set\footnote{Note that the Lyapunov function $V$ of Section \ref{ssir_sec} is the negative of a regularized version of $H$.}
\begin{equation}
H(p,q)=-\beta e^{-p}-\gamma q-\beta e^{-q}.
\end{equation}
It is easy to verify that 
\begin{equation}
\frac{d}{dt}\,H=0,
\end{equation}
under the dynamics \eqref{sirdyn}. The system \eqref{sirdyn} can be explicitly written in the Hamilton form
\begin{equation}
\begin{split}
\dot{p}&=-\frac{\d H}{\d q}\,,\\
\dot{q}&=\frac{\d H}{\d p}\,.
\end{split}
\end{equation}
Notice also that the Hamiltonian function is separable, i.e. it is of the form $H(p,q)=T(p)+V(q)$\footnote{There is a practical dimension to these theoretical observations The fact that the SIR model is a Hamiltonian allows us to solve it numerically by means of the extremely efficient St\"ormer-Verlet method \cite{HLW03}, \cite{LR04}. We will not make use of this method, as it does not seem to directly extend to the stochastic case}. 

A convenient discretization to \eqref{onefactmod} can be formulated in terms of the canonical variables $(p_t,q_t)=(-\log X_{2,t},\,-\log X_{1,t})$ as follows \cite{LL20}. Choose the number of steps $\frak{n}$, define the time step $\delta=T/\frak{n}$, set $t_n=n\delta$ for $n=0,1,\ldots,\frak{n}$, and use simplified $(p_n, q_n)\equiv(p_{t_n},q_{t_n})$. This yields the following Euler scheme:
\begin{equation}
\begin{split}
q_{n+1}&=q_n+(\beta\delta+\sigma \Delta W_n)e^{-p_n}+\frac12\,\sigma^2 \delta e^{-2p_n},\\
p_{n+1}&=p_n+\gamma\delta-(\beta\delta+\sigma \Delta W_n)e^{-q_n}+\frac12\,\sigma^2 \delta e^{-2q_n},\\
\end{split}
\end{equation}
for $n=0,1,\ldots,\frak{n}-1$. Here, the Brownian motion increments $\Delta W=W_{(n+1)\delta}-W_{n\delta}$ are independent variates drawn from the normal distribution $N(0,\delta)$. At each iteration step also have to to floor $q_{n+1}$ and $p_{n+1}$ at $0$, $q_{n+1}=\max(q_{n+1},0)$, $p_{n+1}=\max(p_{n+1},0)$, so that $e^{-q_n}, e^{-p_n}\leq 1$.

Approximating the backward equations of the system \eqref{fbsde_opt} leads to the following backward Euler scheme:
\begin{equation}\label{BSDEEuler}
\begin{split}
Y_{1,n}&=Y_{1,n+1}+f_1(X_n,Y_n,Z_n)\delta-Z_{1,n}\Delta W_n,\\
Y_{2,n}&=Y_{2,n+1}+f_2(X_n,Y_n,Z_n)\delta-Z_{2,n}\Delta W_n,
\end{split}
\end{equation}
where
\begin{equation}\label{gensBSDE}
\begin{split}
f_1(x,y,z)&=\big(\beta(y_2-y_1)+\sigma (z_2-z_1))x_2+N_1-\frac{1}{2 L_1}\,(y_1-M_1)^2,\\
f_2(x,y,z)&=\big(\beta(y_2-y_1)+\sigma (z_2-z_1))x_1-\gamma y_2+N_2-\frac{1}{2 L_2}\,(y_2-M_2)^2
\end{split}
\end{equation}
denote the generators of the two BSDEs. Starting with the terminal condition
\begin{equation}
Y_\frak{n}=\nabla G(X_\frak{n}),
\end{equation}
we will move backward in time with computing $Y_n$ and $Z_n$, for $n=\frak{n}-1,\ldots, 0$.

Notice that two difficulties arise while doing so: (i) the $Y_n$'s in \eqref{BSDEEuler} are not necessarily adapted, and (ii) they depend on $Z_n$. These two problems can be solved by taking conditional expectations $\eE_n(\,\cdot\,)=\eE(\,\cdot\,|\,X_0,X_1,\ldots,X_n)$ on time $t_n$. This leads to the following condition:
\begin{equation*}
\begin{split}
Y_{i,n}&= \eE_n(Y_{i,n}) \\
&=\eE_n(Y_{i,n+1})+ f_i(X_n,Y_n, Z_n)\delta.
\end{split}
\end{equation*}
This is an implicit scheme, which may slow down the computations. However, we can easily replace it with an explicit scheme of the same order: 
\begin{equation*}
\begin{split}
Y_{i,n}&=\eE_n\big(Y_{i,n+1}+f_i(X_n,Y_{i,n+1}, Z_n)\delta\big).
\end{split}
\end{equation*}

In order to determine $Z_{i,n}$, $i=1,2$, we multiply \eqref{BSDEEuler} by an increment $\Delta W_n$ and take conditional expectations. This yields
\begin{equation*}
\begin{split}
0 &= \eE_n(Y_{i,n} \Delta W_n)\\
&= \eE_n(Y_{i,n+1} \Delta W_n)- Z_{i,n}\delta,
\end{split}
\end{equation*}
and hence we obtain the following expression for $Z_{i,n}$:
\begin{equation*}
Z_{i,n}=\frac{1}{\delta}\,\eE_n(Y_{i,n+1} \Delta W_n).
\end{equation*}

In summary, we are led to the following discrete time scheme for solving \eqref{BSDEEuler}:
\begin{equation}\label{discrFbSystem}
\begin{split}
Y_\frak{n} &=\nabla G(X_\frak{n}),\\
Z_{i,n}&=\frac{1}{\delta}\,\eE_n(Y_{i,n+1} \Delta W_n),\\
Y_{i,n}&=\eE_n\big(Y_{i,n+1}+f_i(X_n,Y_{n+1}, Z_n)\delta\big),\\
\end{split}
\end{equation}
for $n=\frak{n}-1,\ldots, 0$. Note that simulating this system requires numerical estimation of the conditional expected values $\eE_n(\,\cdot \,)$ in the formulas above. We discuss this issue in the following section. 

\subsection{Computing the conditional expectations}
\label{subsec:HermitePolynomials}

A practical and powerful method of computing the conditional expected values in \eqref{discrFbSystem} is the Longstaff-Schwartz regression method originally developed for modeling American options \cite{LS01}. We use a variant of this method that involves the Hermite polynomials, and that was used for a similar purpose in \cite{LR16}. This choice is natural, as conditional expectations of Hermite polynomials of a Gaussian random variable lead to simple closed form expressions. 

Let $\mathrm{He}_k\ofx$, $k=0,1,\ldots$, denote the $k$-th normalized Hermite polynomial corresponding to the standard Gaussian measure $d\mu(x)=(2\pi)^{-1/2}\,e^{-x^2/2} dx$. These functions form an orthonormal basis for the Hilbert space $L^2 \big(\bR, d\mu\big)$.

The key property of $\mathrm{He}_k(x)$ is the following addition formula for $\chi \in [0,1]$ and $w,x \in \bR$:
\begin{equation} \label{HeAdd}
\mathrm{He}_k(\sqrt{\chi}w+\sqrt{1-\chi}x)
=\sum_{j=0}^k{k\choose j}^{1/2}\,\chi^{j/2}(1-\chi)^{(k-j)/2}\mathrm{He}_j(w)\mathrm{He}_{k-j}(x).
\end{equation}
Consequently, integrating over $x$ with respect to to the measure $\mu$ yields the following conditioning rule:
\begin{equation}\label{condRule}
\eE\big(\mathrm{He}_k(\sqrt{\chi}w+\sqrt{1-\chi}x)\,|\,w\big)=\chi^{k/2}\mathrm{He}_k(w).
\end{equation}
Here, $w, x$ are independent standard normal random variables. We shall use this relation in order to estimate the conditional expected values in \eqref{discrFbSystem}.

We set $W_{t_i}=\sqrt{t_i}\,w_i$, for $i=1,\ldots, m$, where $w_i$ is an $n$-dimensional standard normal random variable. We notice that
\begin{equation}\label{orthDecomp}
w_{i+1}=\sqrt{\chi_i}\,w_i+\sqrt{1-\chi_i}\,X_i,
\end{equation}
where $\chi_i=t_i/t_{i+1}$, and where $X_i$ is standard normal and independent of $w_i$. In the following, we shall use this decomposition in conjunction with \eqref{condRule}.

Now, we assume the following linear architecture:
\begin{equation}\label{hermExp}
Y_{i+1}=\sum_{k=0}^K\,g_{k,i+1}\mathrm{He}_k(w_{i+1}),
\end{equation}
where $K$ is the cutoff value of the order of the Hermite polynomials. This is simply a truncated expansion of the random variable $Y_{i+1}$ in terms of the orthonormal basis $\mathrm{He}_k(w_{i+1})$. The values of the Fourier coefficients are \textit{estimated by means of ordinary least square regression}. Then, as a consequence of the conditioning rule \eqref{condRule},
\begin{equation}\label{condExp}
\eE_i(Y_{i+1})=\sum_{k=0}^K g_{k,i+1}\,\chi_i^{k/2}\,\mathrm{He}_k(w_i).
\end{equation}
In other words, conditioning $Y_{i+1}$ on $w_{i}$ is equivalent to multiplying its Fourier coefficients $g_k$ by the factor $\chi_i^{k/2}$. This allows us to calculate the first term on the right hand side of the third equation in \eqref{discrFbSystem}. In order to calculate the second term, we substitute the explicit formula for $Z_n$ into the generators \eqref{gensBSDE} and repeat the calculations in \eqref{hermExp} and \eqref{condExp} with $Y_{n+1}$ replaced by $Y_{n+1} + f(X_n,Y_n,Z_n)\delta$.

\subsection{Numerical experiments}

\begin{figure}[H]
\scalebox{1.0}[1.0]{\includegraphics[height=6.6cm]{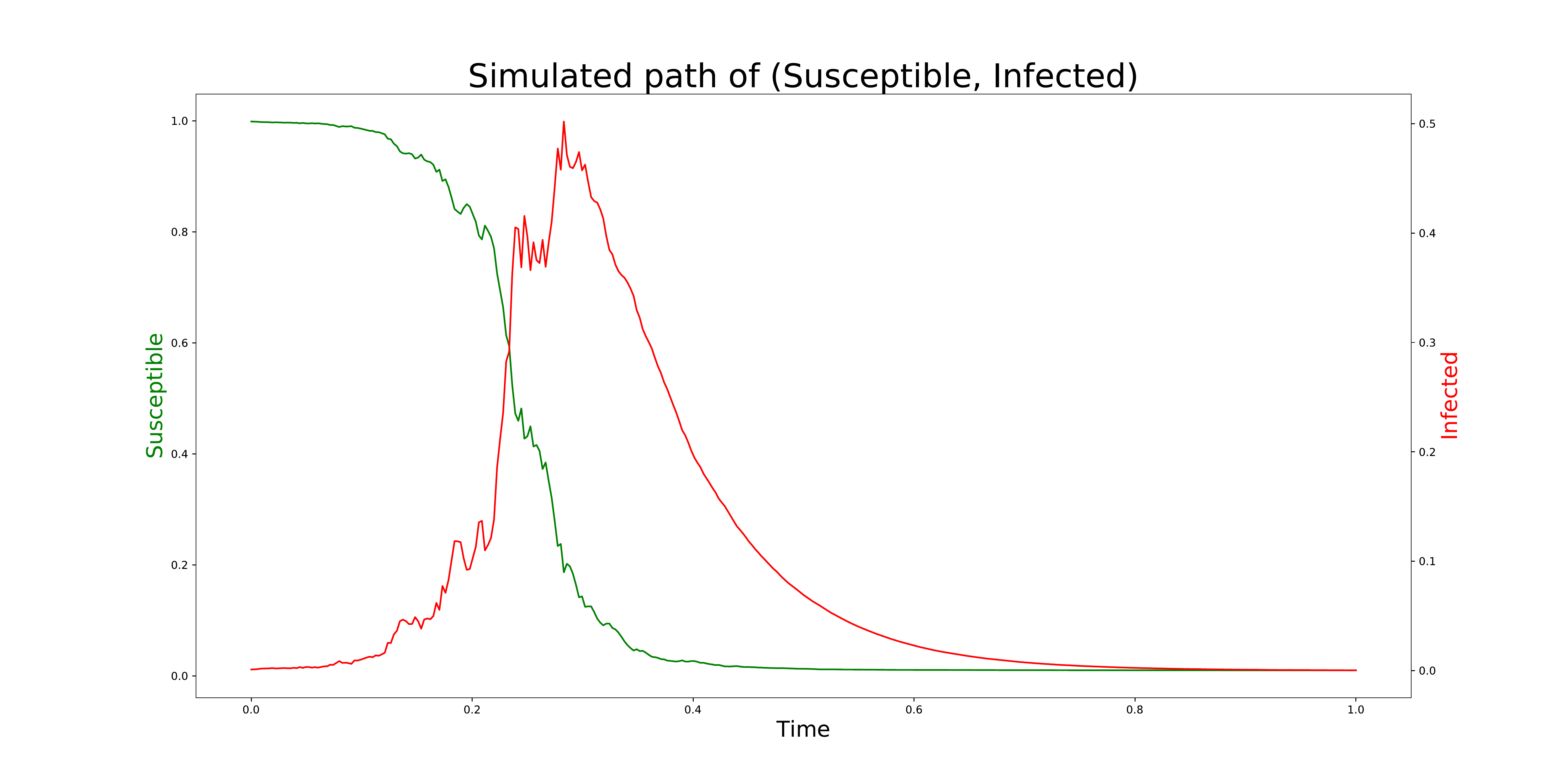}}
\caption{\label{SI_raw} A single Monte Carlo path of the raw process \eqref{onefactmod}.}
\end{figure}

For numerical experiments within the framework developed above, we assume a time horizon of 1 year (365 days), and choose the SIR model parameters as follows:
\begin{equation}
\begin{split}
\beta&=38.0,\\
\gamma&=11.5,\\
S_0&=0.999,\\
I_0&=0.001,
\end{split}
\end{equation}
These parameters are purely hypothetical, and they do not arise from the calibration to any actually reported data. The corresponding value of the basic reproduction ratio $R_0=S_0\beta/\gamma$ is 3.30 and it indicates a highly infectious disease such as COVID-19. We choose the diffusion parameter
\begin{equation}
\sigma=3.1.
\end{equation}
A typical scenario generated by this model is graphed in Figure \ref{SI_raw}.

For solving the FBSDE system \eqref{fbsde_opt} we generate $2,000$ scenarios (Monte Carlo paths) using the low variance spectral decomposition method. Each scenario is based on daily sampling, i.e. $\frak{n}=365$. For calculating the conditional expectations in the Longstaff-Schwartz regression we use the cutoff value of $K=4$. The expectation behind this choice is that we obtain the accuracy of four sigma.

We attempt to construct a numerical solution to \eqref{fbsde_opt} through the following iterative procedure. We start by generating initial Monte Carlo paths $X_t^{(0)}$ using the (discretized version of the) raw process \eqref{onefactmod}. For the initial guess of $Y_t$ we take $Y_t^{(0)}=\nabla G(X_t^{(0)})$. Notice that this is \textit{not} at solution to the backward equations in \eqref{fbsde_opt}, it merely satisfies the terminal condition. After this, we iterate
\begin{equation}
\begin{split}
X_t^{(k+1)}&=\text{ solution of the discretized forward equation with $Y_t=Y_t^{(k)}$,}\\
Y_t^{(k+1)}&=\text{ solution of the discretized backward equation with $X_t=X_t^{(k+1)}$,}
\end{split}
\end{equation}
for $k=0,1,\ldots$, until the stopping criterion is satisfied or the maximum number of iterations is reached. For the stopping criterion we choose the condition that the average $L^2$-norm change of a Monte Carlo path falls below a tolerance threshold of $10^{-8}$.

We make various choices of the coefficients $L,M,N$ defining the running cost functions. Depending on the values of these coefficients, the iterative process described above converges to a meaningful solution or it diverges. At this point, it is unclear what choices of the coefficients lead to what outcomes, but it appears that there are well defined basins of convergence in the space of the parameters.

\begin{figure}[H]
\scalebox{1.0}[1.0]{\includegraphics[height=6.6cm]{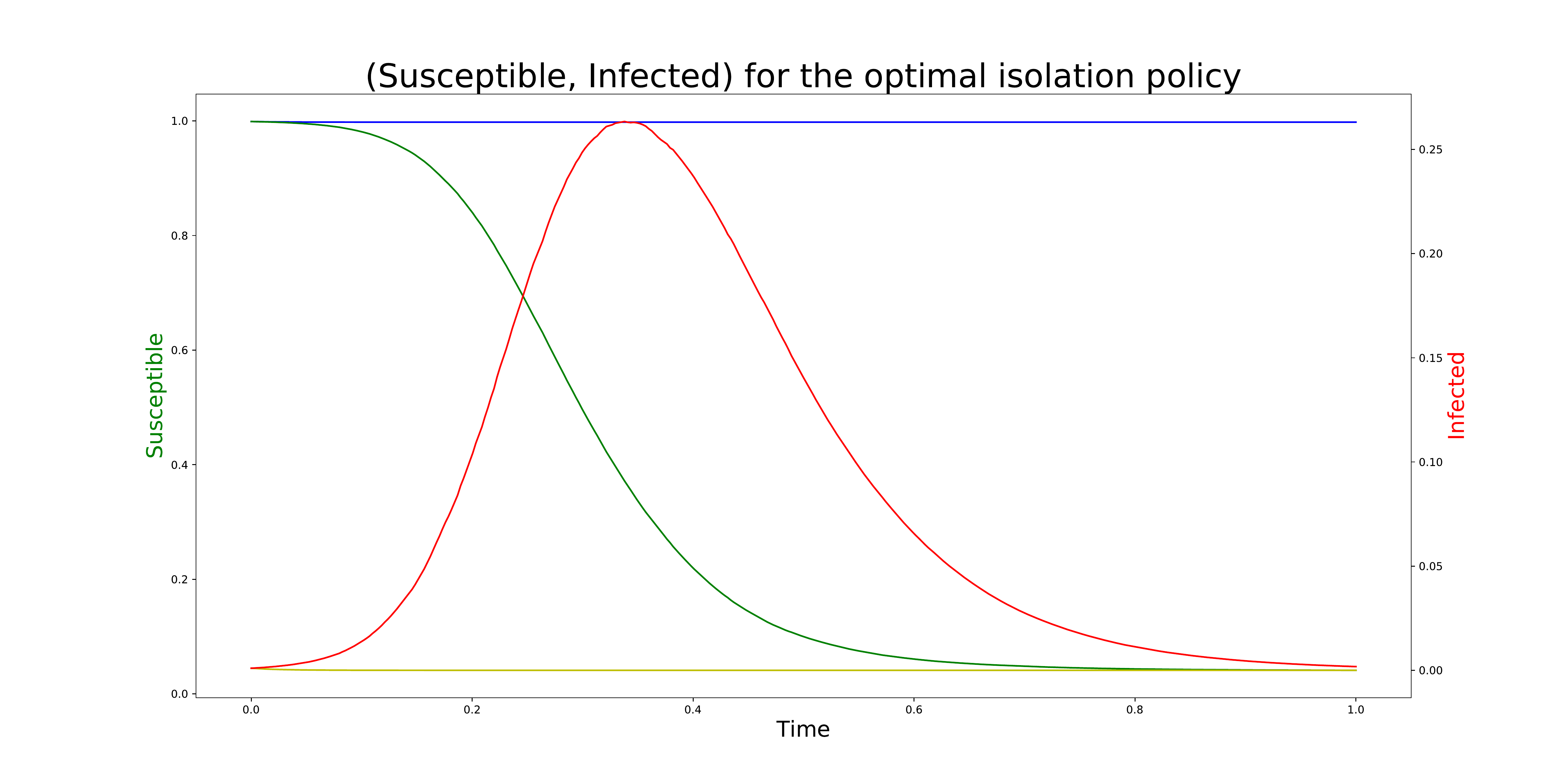}}
\caption{\label{SI_iso_1}\small Raw (Susceptible, Infected) versus optimal (Susceptible, Infected) under the isolation policy assuming high cost running cost function. The read and green lines are the average values of the raw Susceptible (primary axis) and Infected (secondary axis) fractions, respectively, while the yellow and blue lines represent the averages of the corresponding optimal values.}
\end{figure}

\begin{figure}[H]
\scalebox{1.0}[1.0]{\includegraphics[height=6.6cm]{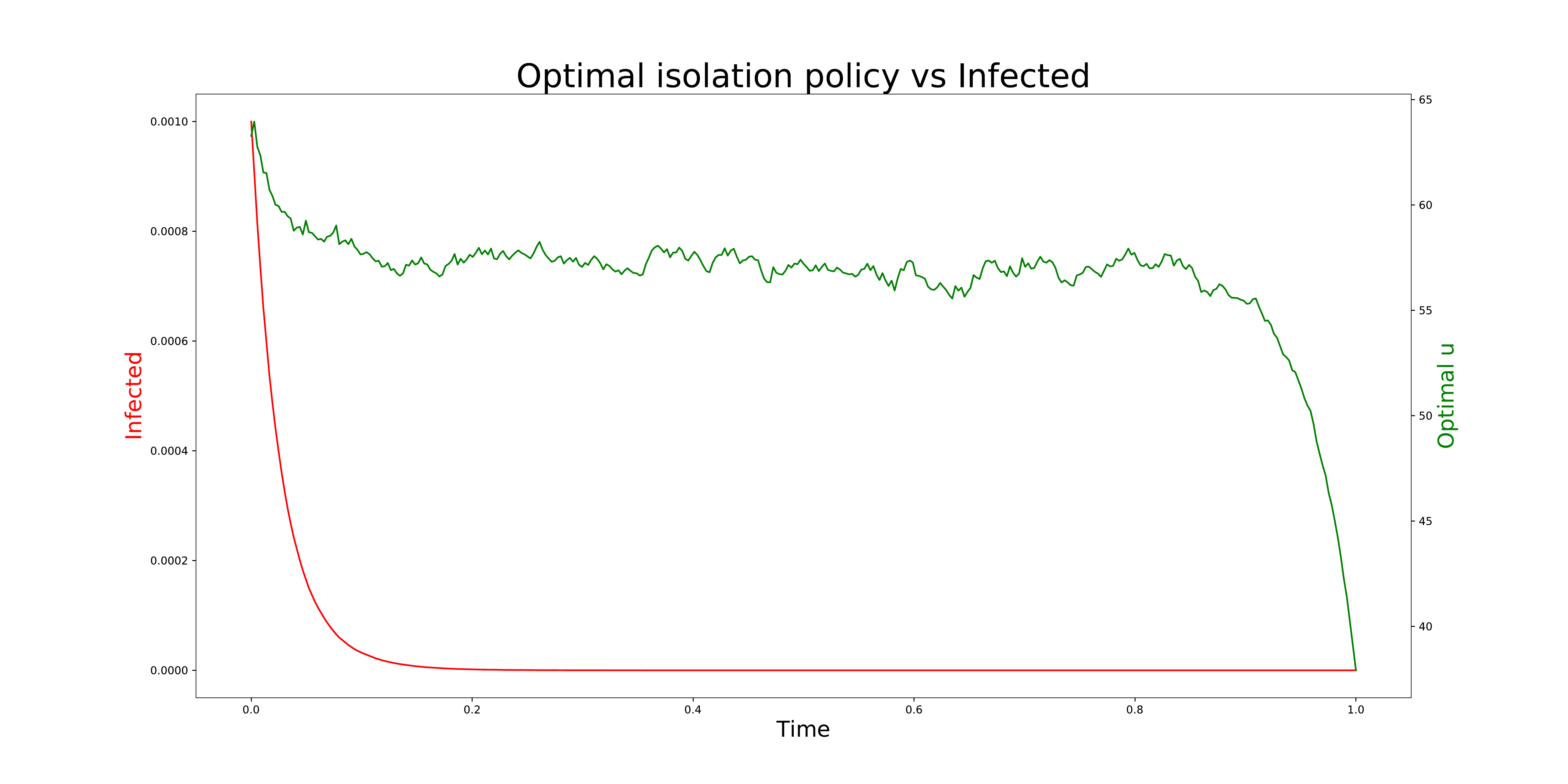}}
\caption{\label{SI_iso_vs_I_1}\small Plot of the Infected fraction (primary axis) versus the optimal isolation policy (secondary axis) under high cost running cost function.}
\end{figure}

We first consider a high cost policy. Figures \ref{SI_iso_1} - \ref{SI_iso_vs_I_1} show the graphs of the solutions assuming the following parameters of the quadratic polynomial in the running cost function $c_2(u_2,x_2)$:
\begin{equation}
\begin{split}
L_2&=1.0,\\
M_2&=0.0,\\
N_2&=120.0.
\end{split}
\end{equation}
Unlike Figure \ref{SI_raw}, the curves in all graphs below show the averages of the corresponding quantities over 2,000 Monte Carlo paths.
\begin{figure}[H]
\scalebox{1.0}[1.0]{\includegraphics[height=6.6cm]{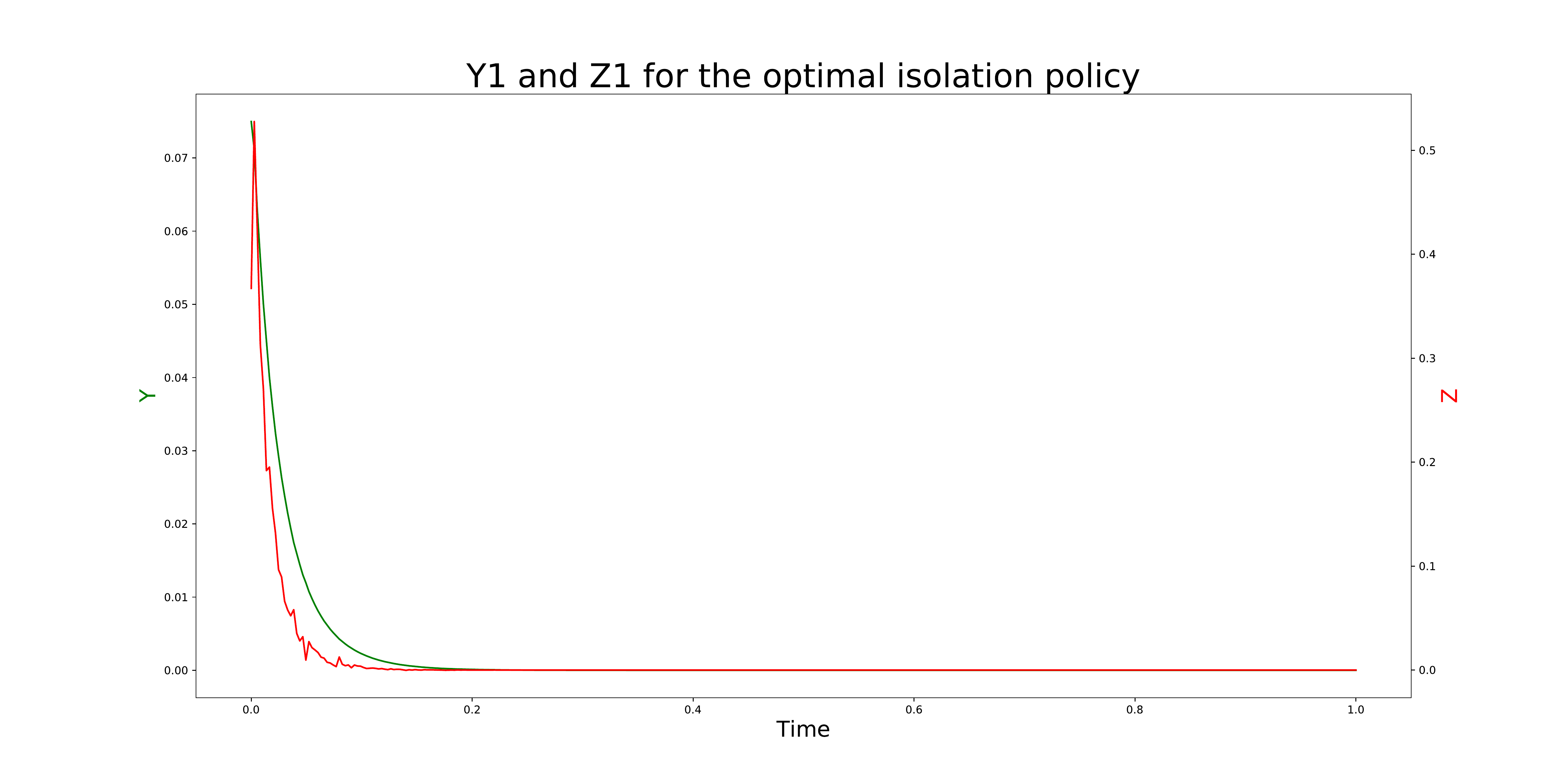}}
\caption{\label{SI_Y1Z1iso_1}\small Plot of $Y_{1,t}$ (primary axis) and $Z_{1,t}$ (secondary axis) for equation \eqref{fbsde_opt_i}.}
\end{figure}
\begin{figure}[H]
\scalebox{1.0}[1.0]{\includegraphics[height=6.6cm]{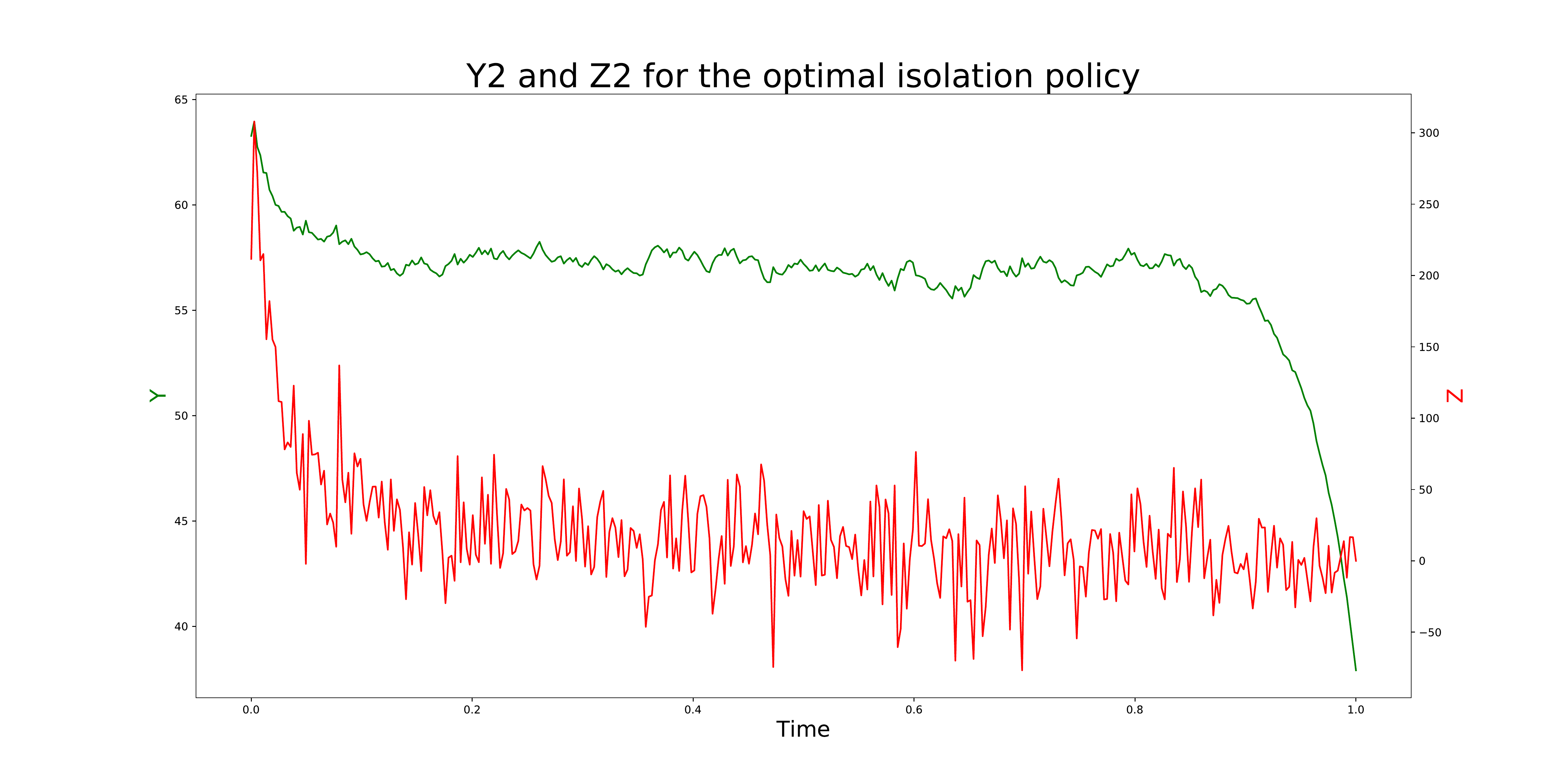}}
\caption{\label{SI_Y1Z1iso_1}\small Plot of $Y_{2,t}$ (primary axis) and $Z_{2,t}$ (secondary axis) for equation \eqref{fbsde_opt_i}.}
\end{figure}

Under this running cost function, the optimal policy is to implement a draconian isolation regime, which leads to a rapid drop in infections, while keeping the susceptible fraction of the population at a very high level.

\begin{figure}[H]
\scalebox{1.0}[1.0]{\includegraphics[height=6.6cm]{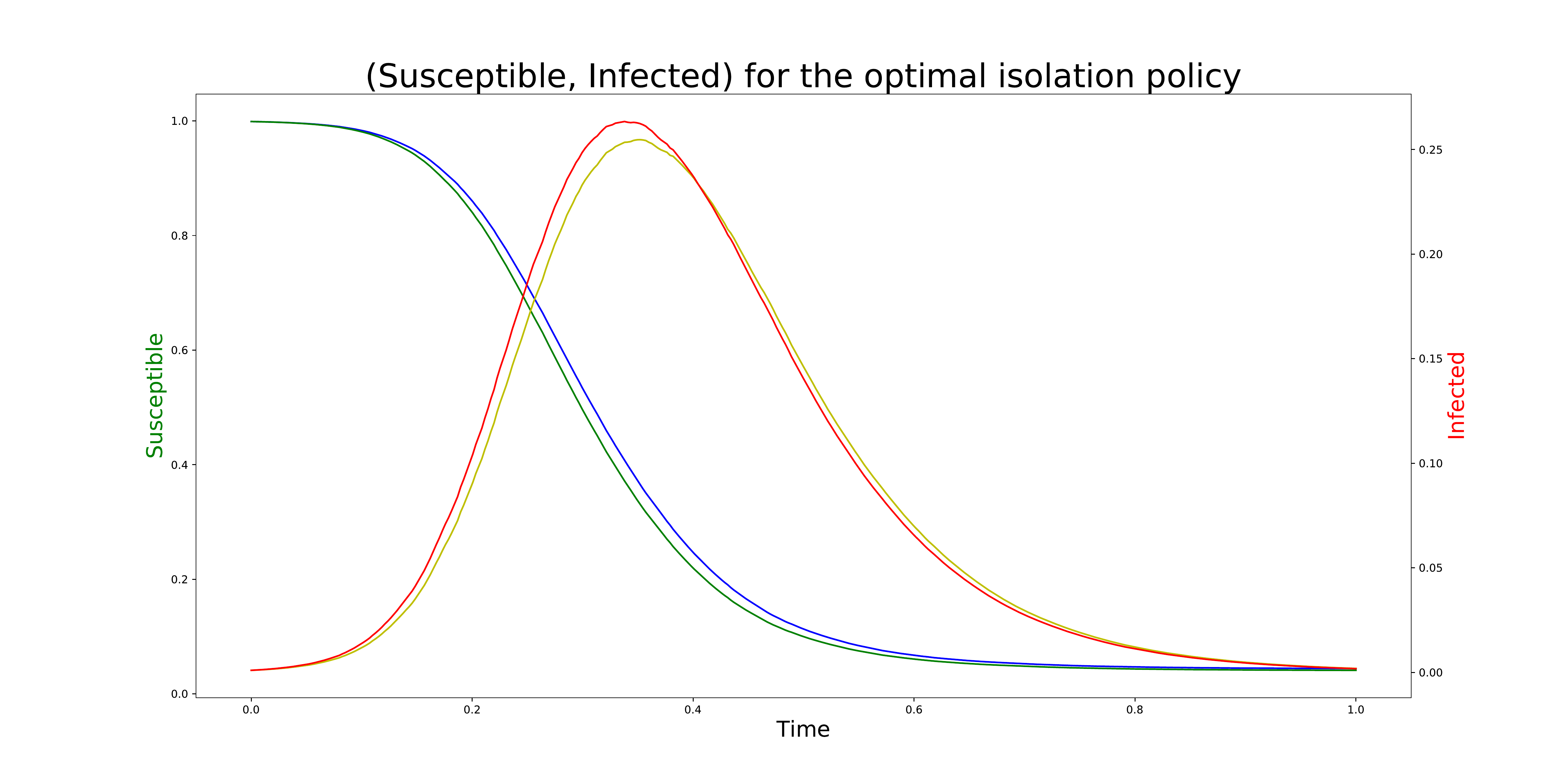}}
\caption{\label{SI_iso_2} \small Raw (Susceptible, Infected) versus optimal (Susceptible, Infected) under the isolation policy assuming low cost running cost function. The read and green lines are the average values of the raw Susceptible (primary axis) and Infected (secondary axis) fractions, respectively, while the yellow and blue lines represent the averages of the corresponding optimal values.}
\end{figure}

\begin{figure}[H]
\scalebox{1.0}[1.0]{\includegraphics[height=6.6cm]{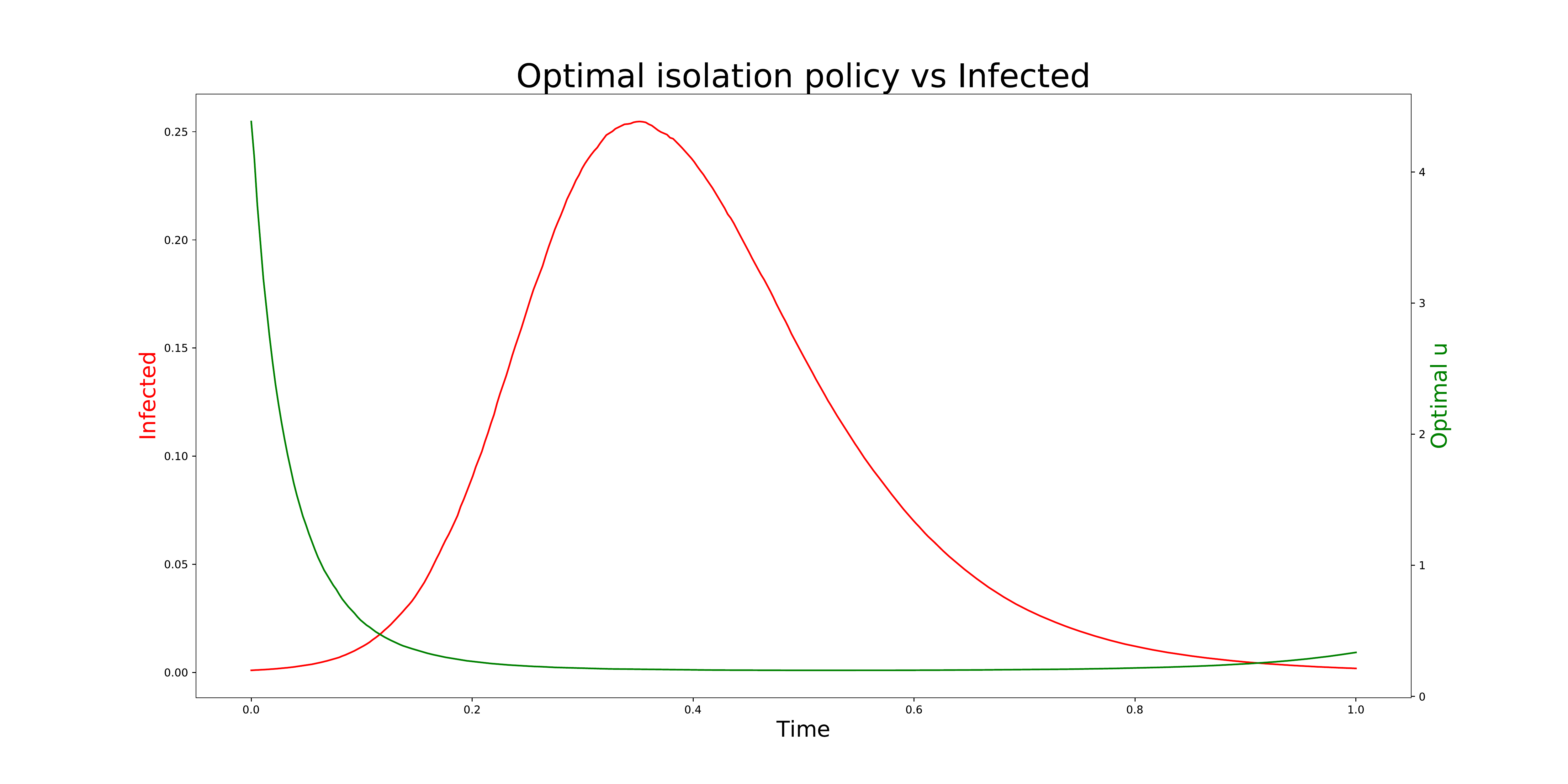}}
\caption{\label{SI_iso_vs_I_2}\small Plot of the Infected fraction (primary axis) versus the optimal isolation policy (secondary axis) under low cost running cost function.}
\end{figure}

On the other hand, Figures \ref{SI_iso_2} - \ref{SI_iso_vs_I_2} show the plots of the solutions assuming a low cost policy, with the following parameters of the quadratic polynomial in the running cost function $c_2(u_2,x_2)$:
\begin{equation}
\begin{split}
L_2&=5.0,\\
M_2&=0.0,\\
N_2&=10.0.
\end{split}
\end{equation}
Under this running cost function, the optimal policy is a moderate isolation regime. Following this policy, the isolation rate is high, as the infections are low, and then it declines over time while the epidemic develops. Unlike the policy above, this leads to a gradual decline in both the infected and susceptible fractions of the population.
\begin{figure}[H]
\scalebox{1.0}[1.0]{\includegraphics[height=6.6cm]{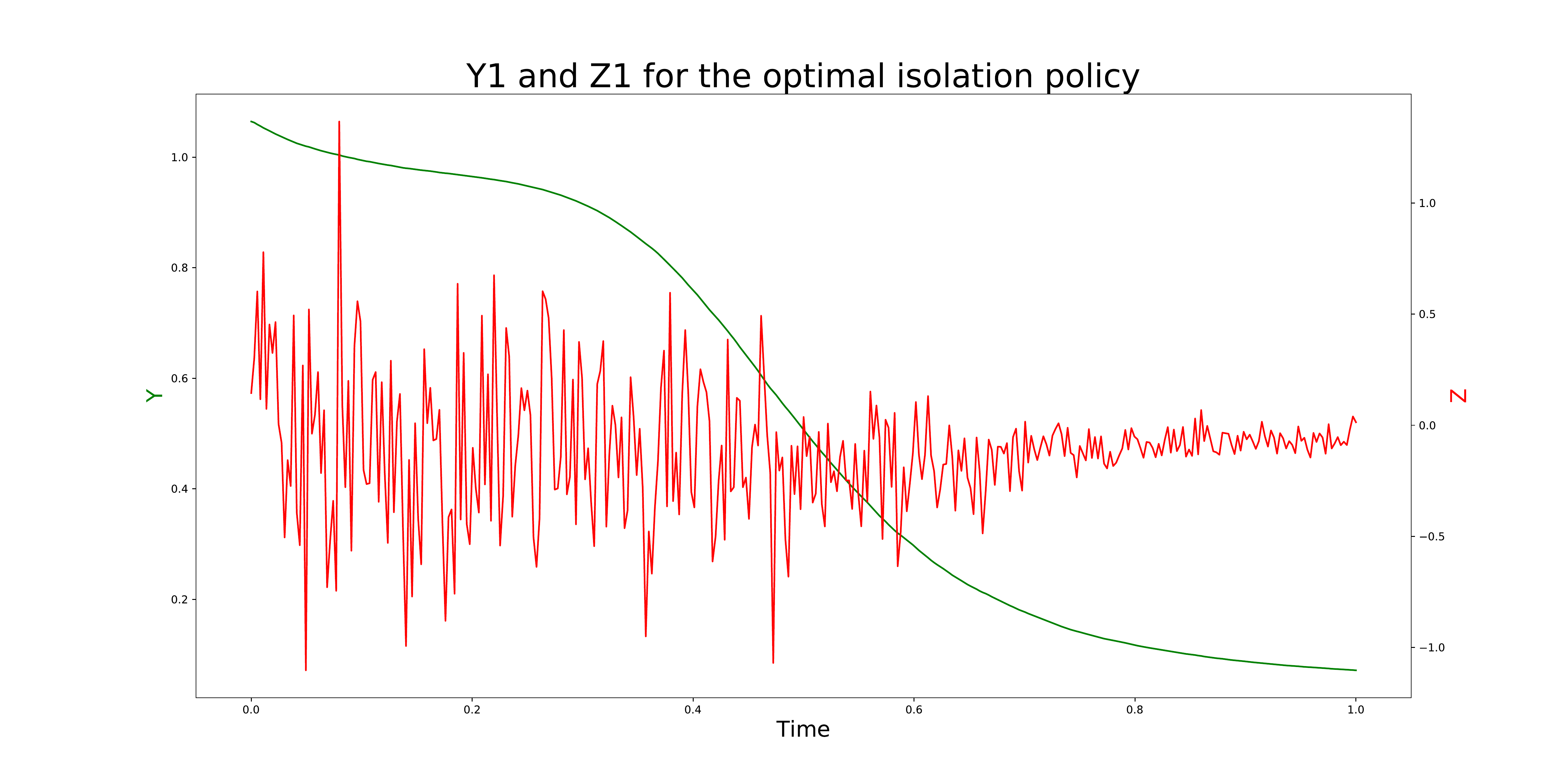}}
\caption{\label{SI_Y1Z1iso_2}\small Plot of $Y_{1,t}$ (primary axis) and $Z_{1,t}$ (secondary axis) for equation \eqref{fbsde_opt_i}.}
\end{figure}
\begin{figure}[H]
\scalebox{1.0}[1.0]{\includegraphics[height=6.6cm]{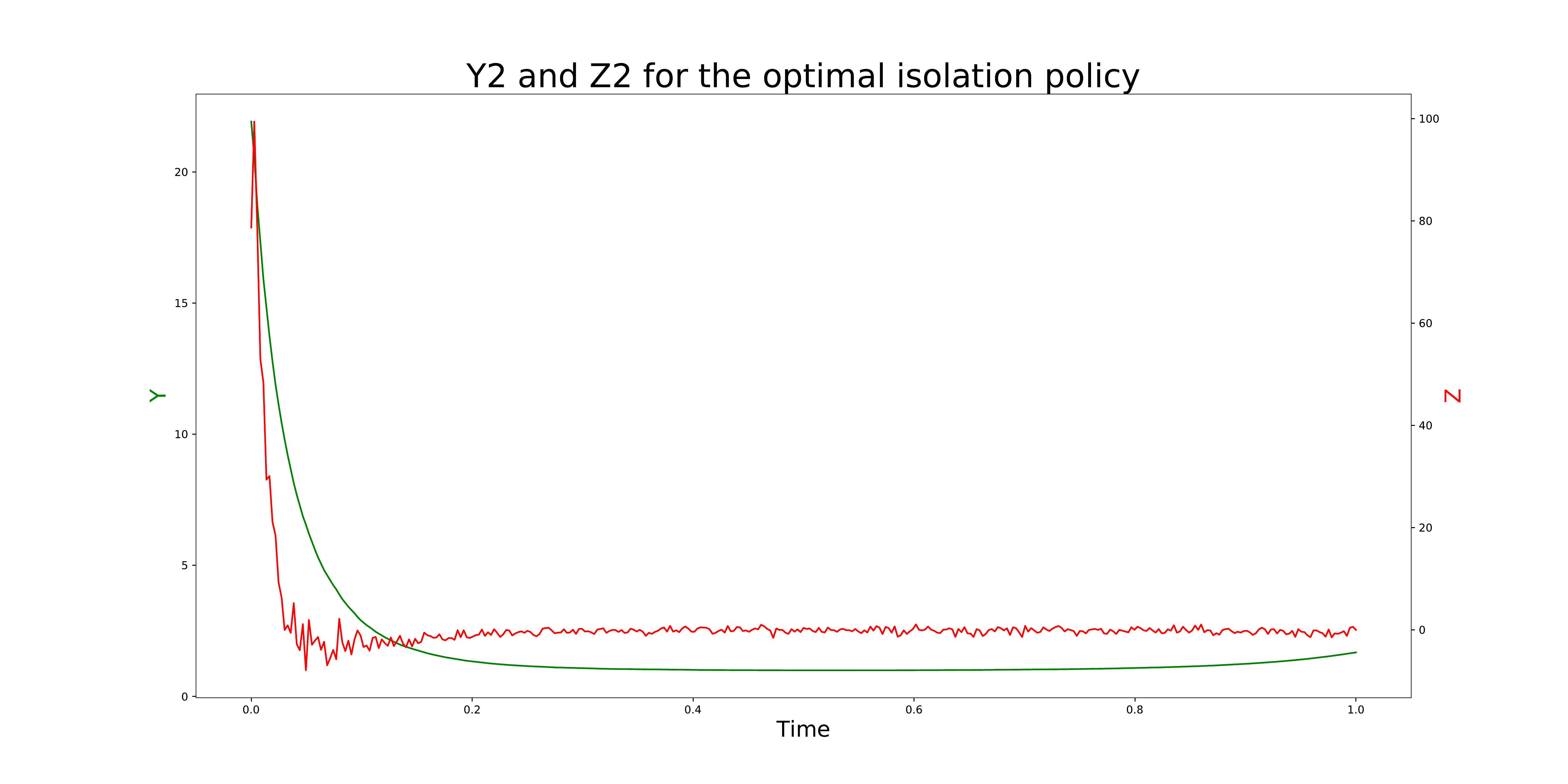}}
\caption{\label{SI_Y1Z1iso_2}\small Plot of $Y_{2,t}$ (primary axis) and $Z_{2,t}$ (secondary axis) for equation \eqref{fbsde_opt_i}.}
\end{figure}

Consider now the case of an optimal vaccination strategy. Again, we make two choices of the running cost function: high cost and low cost. 

\begin{figure}[H]
\scalebox{1.0}[1.0]{\includegraphics[height=6.6cm]{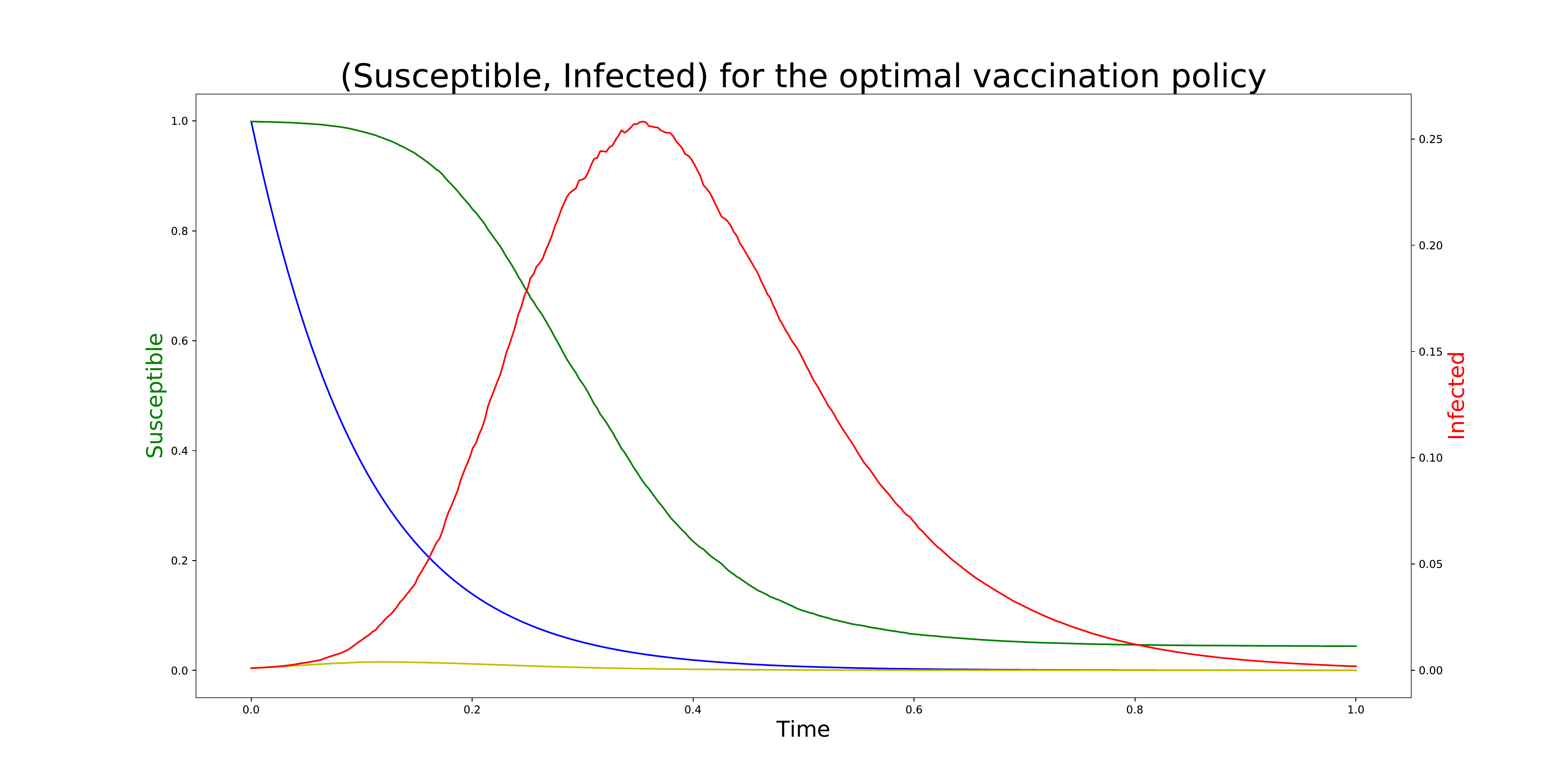}}
\caption{\label{SI_vac_1}\small Raw (Susceptible, Infected) versus optimal (Susceptible, Infected) under the vaccination policy assuming high cost running cost function. The read and green lines are the average values of the raw Susceptible (primary axis) and Infected (secondary axis) fractions, respectively, while the yellow and blue lines represent the averages of the corresponding optimal values.}
\end{figure}

\begin{figure}[H]
\scalebox{1.0}[1.0]{\includegraphics[height=6.6cm]{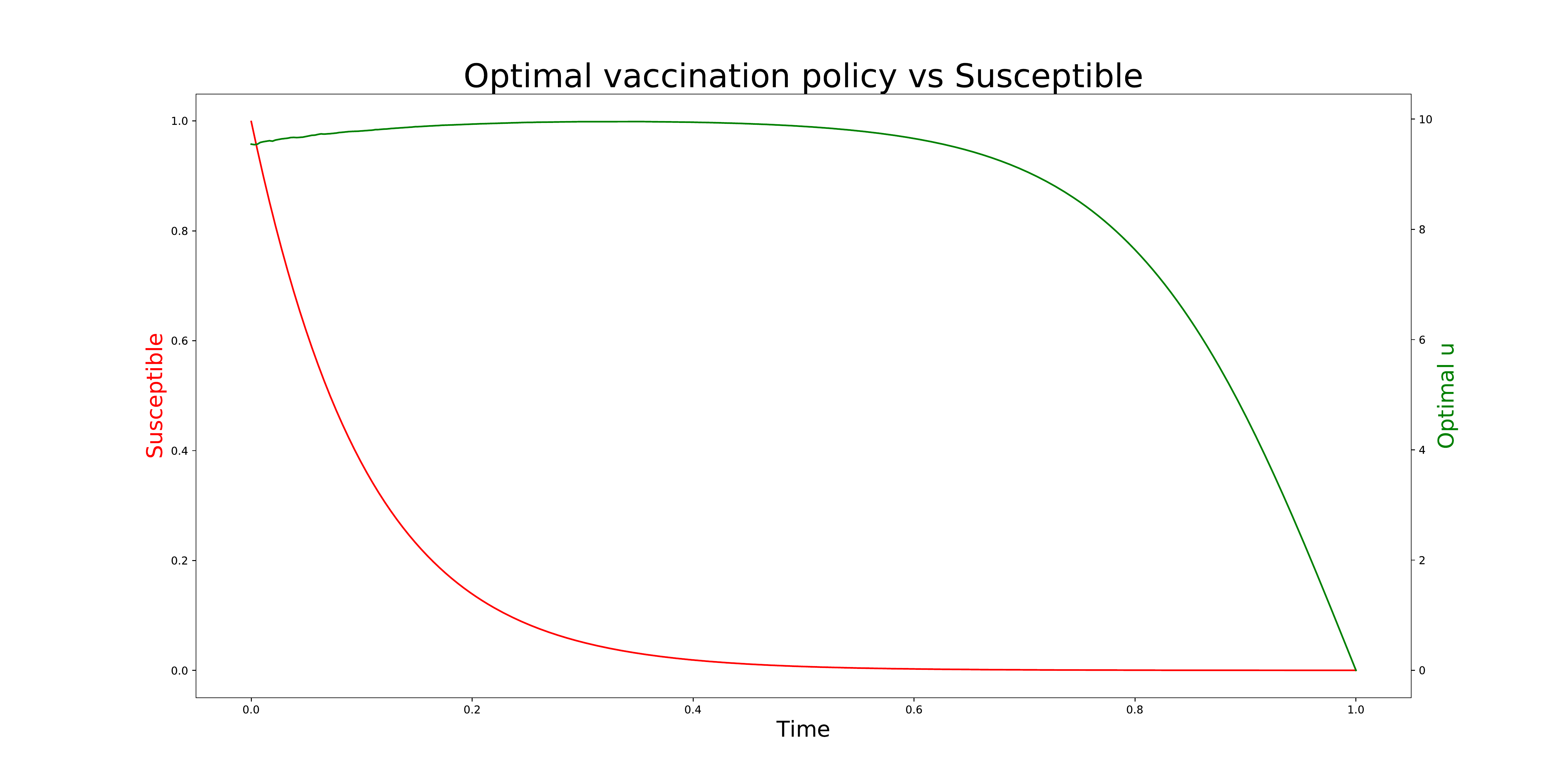}}
\caption{\label{SI_vac_vs_S_1}\small Plot of the Infected fraction (primary axis) versus the optimal vaccination policy (secondary axis) under high cost running cost function.}
\end{figure}

For the high cost case we choose the parameters as follows:
\begin{equation}
\begin{split}
L_1&=10.0,\\
M_1&=0.0,\\
N_1&=500.0.
\end{split}
\end{equation}
The results of Monte Carlo simulations for this cost function are plotted in Figures \ref{SI_vac_1} - \ref{SI_vac_vs_S_1}. They parallel the results presented above in the case of high cost isolation mitigation. The optimal policy is a massive vaccination campaign that dramatically reduces the susceptible fraction of the population and leads to significantly lower infections. 
\begin{figure}[H]
\scalebox{1.0}[1.0]{\includegraphics[height=6.6cm]{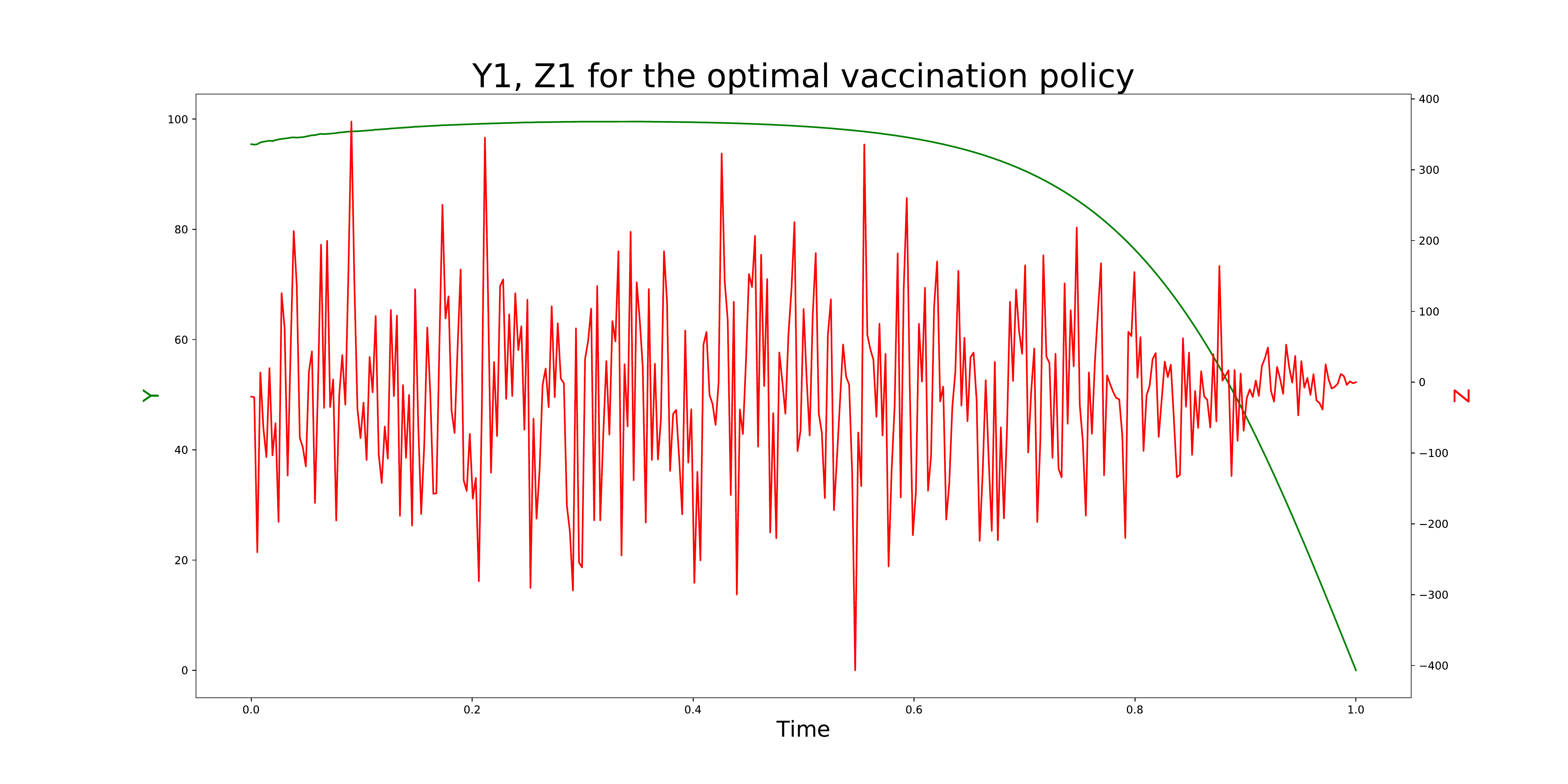}}
\caption{\label{SI_Y1Z1vac_1}\small Plot of $Y_{1,t}$ (primary axis) and $Z_{1,t}$ (secondary axis) for equation \eqref{fbsde_opt_v}.}
\end{figure}
\begin{figure}[H]
\scalebox{1.0}[1.0]{\includegraphics[height=6.6cm]{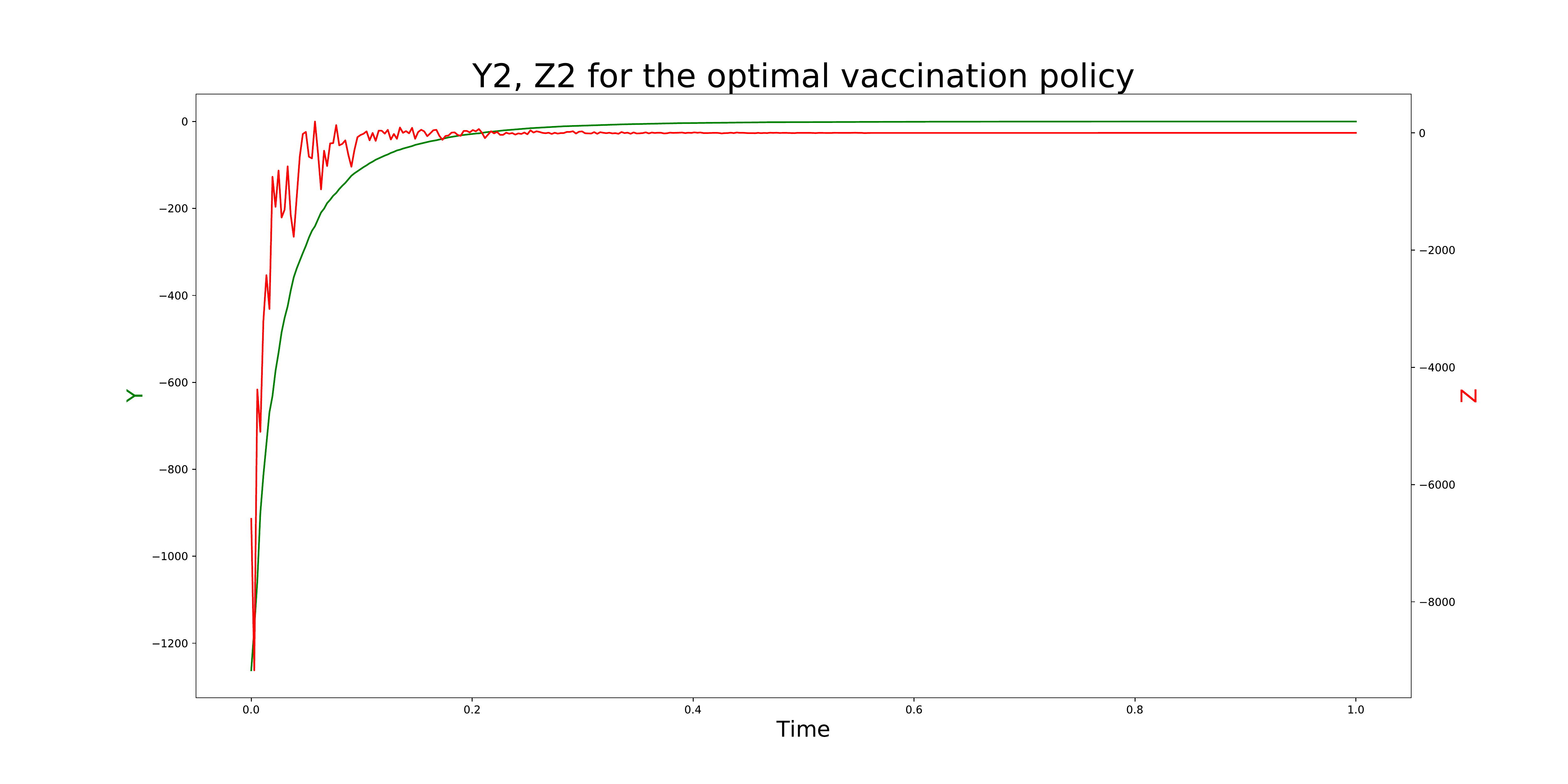}}
\caption{\label{SI_Y1Z1vac_1}\small Plot of $Y_{2,t}$ (primary axis) and $Z_{2,t}$ (secondary axis) for equation \eqref{fbsde_opt_v}.}
\end{figure}

\begin{figure}[H]
\scalebox{1.0}[1.0]{\includegraphics[height=6.6cm]{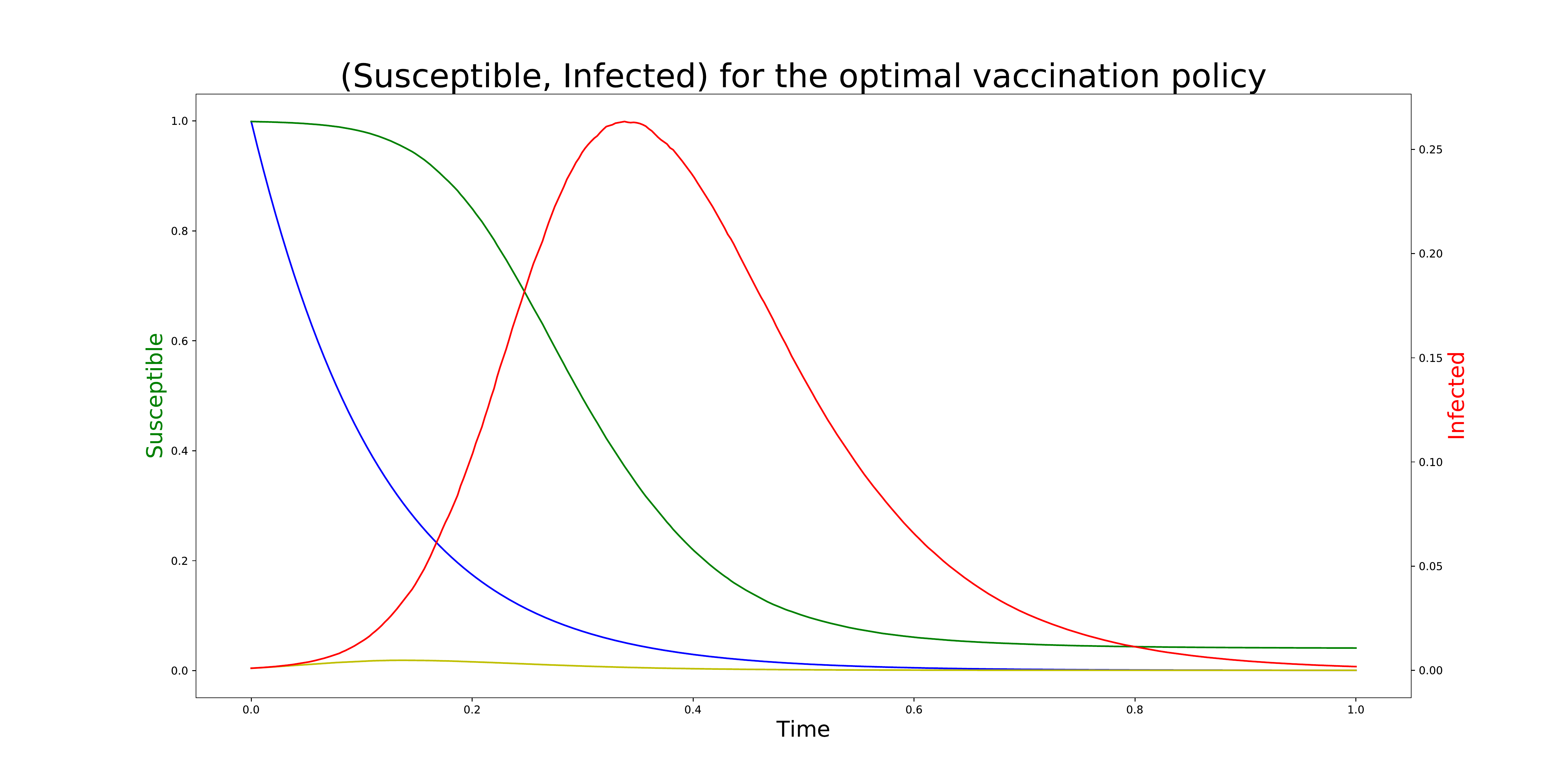}}
\caption{\label{SI_vac_2}\small Raw (Susceptible, Infected) versus optimal (Susceptible, Infected) under the vaccination policy assuming low cost running cost function. The read and green lines are the average values of the raw Susceptible (primary axis) and Infected (secondary axis) fractions, respectively, while the yellow and blue lines represent the averages of the corresponding optimal values.}
\end{figure}

\begin{figure}[H]
\scalebox{1.0}[1.0]{\includegraphics[height=6.6cm]{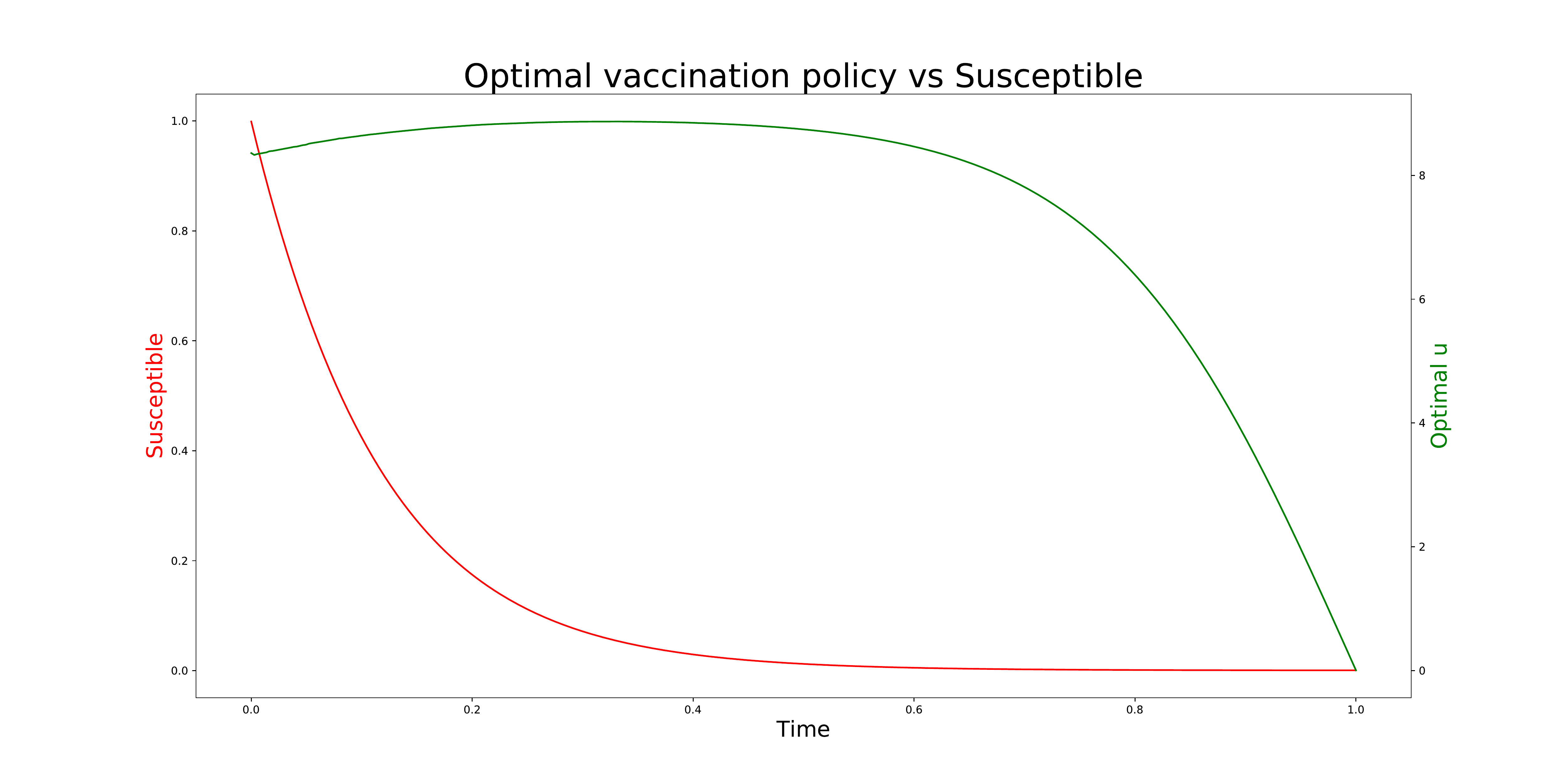}}
\caption{\label{SI_vac_vs_S_2}\small Plot of the Infected fraction (primary axis) versus the optimal vaccination policy (secondary axis) under low cost running cost function.}
\end{figure}

The low cost running cost function is parameterized as follows:
\begin{equation}
\begin{split}
L_2&=0.01,\\
M_2&=0.0,\\
N_2&=0.4.
\end{split}
\end{equation}
The results of Monte Carlo simulations for this cost function are plotted in Figures \ref{SI_vac_2} - \ref{SI_vac_vs_S_2}.
\begin{figure}[H]
\scalebox{1.0}[1.0]{\includegraphics[height=6.6cm]{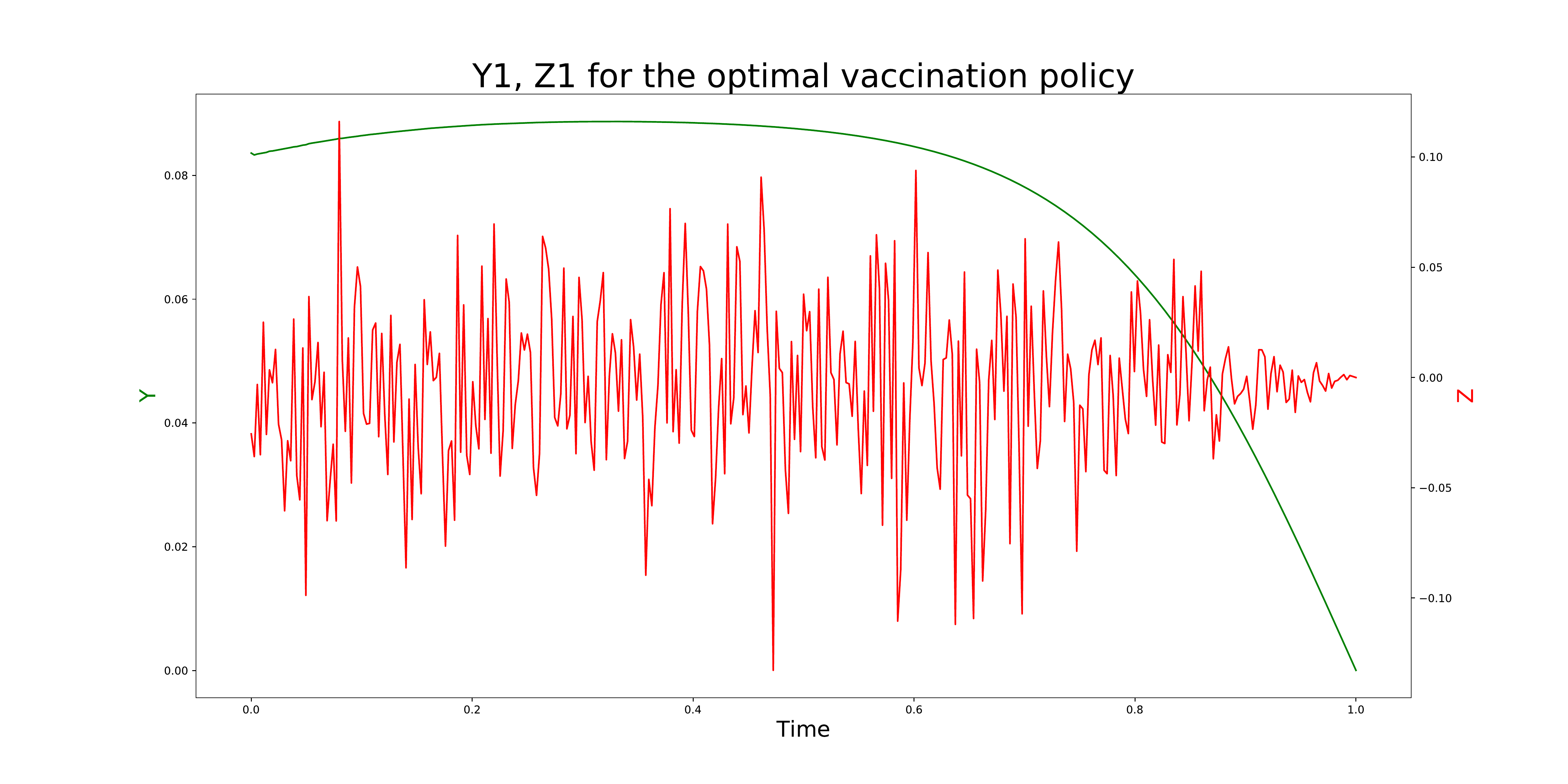}}
\caption{\label{SI_Y1Z1vac_2}\small Plot of $Y_{1,t}$ (primary axis) and $Z_{1,t}$ (secondary axis) for equation \eqref{fbsde_opt_v}.}
\end{figure}
\begin{figure}[H]
\scalebox{1.0}[1.0]{\includegraphics[height=6.6cm]{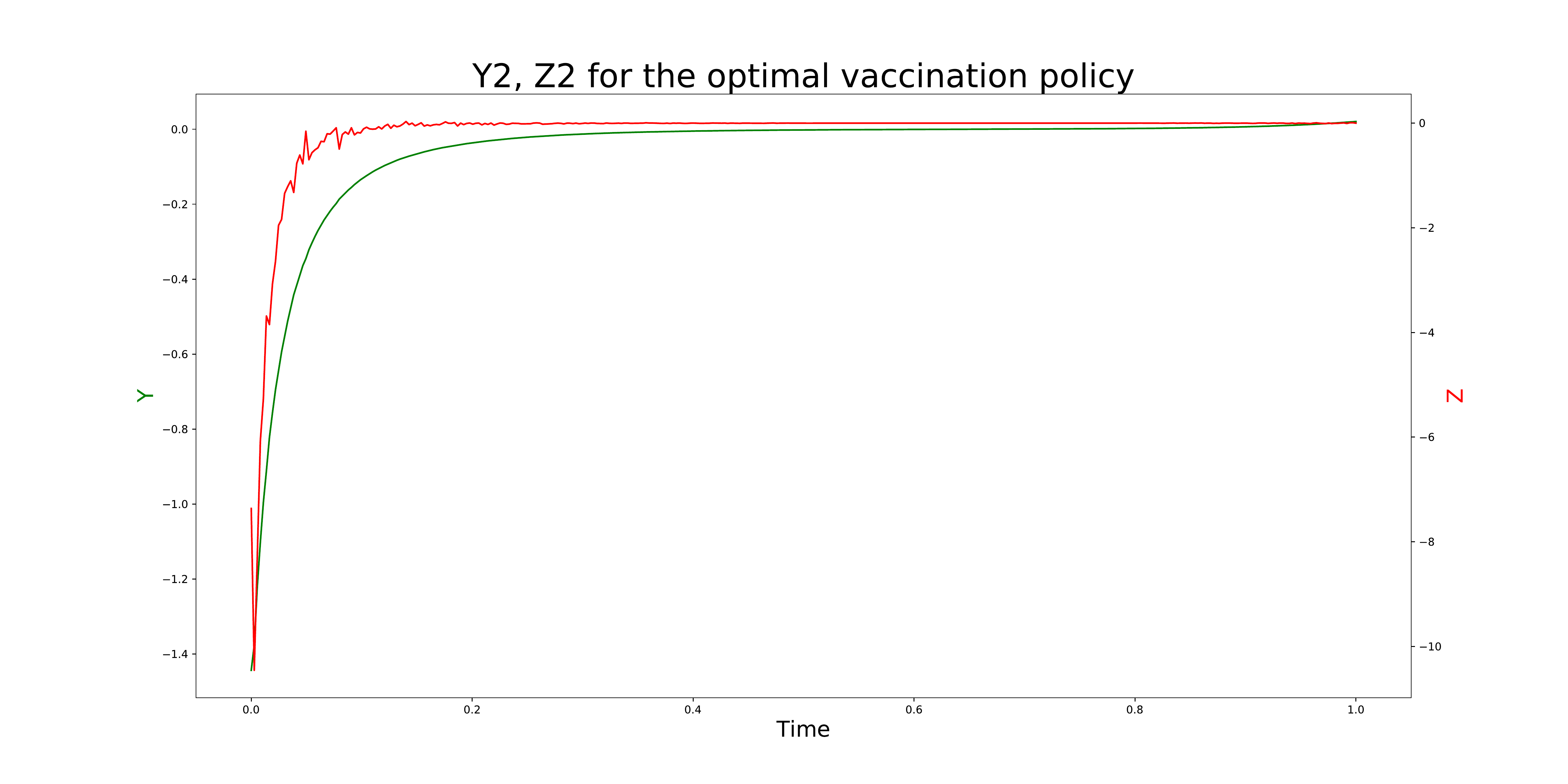}}
\caption{\label{SI_Y1Z1vac_2}\small Plot of $Y_{2,t}$ (primary axis) and $Z_{2,t}$ (secondary axis) for equation \eqref{fbsde_opt_v}.}
\end{figure}

\begin{figure}[H]
\scalebox{1.0}[1.0]{\includegraphics[height=6.6cm]{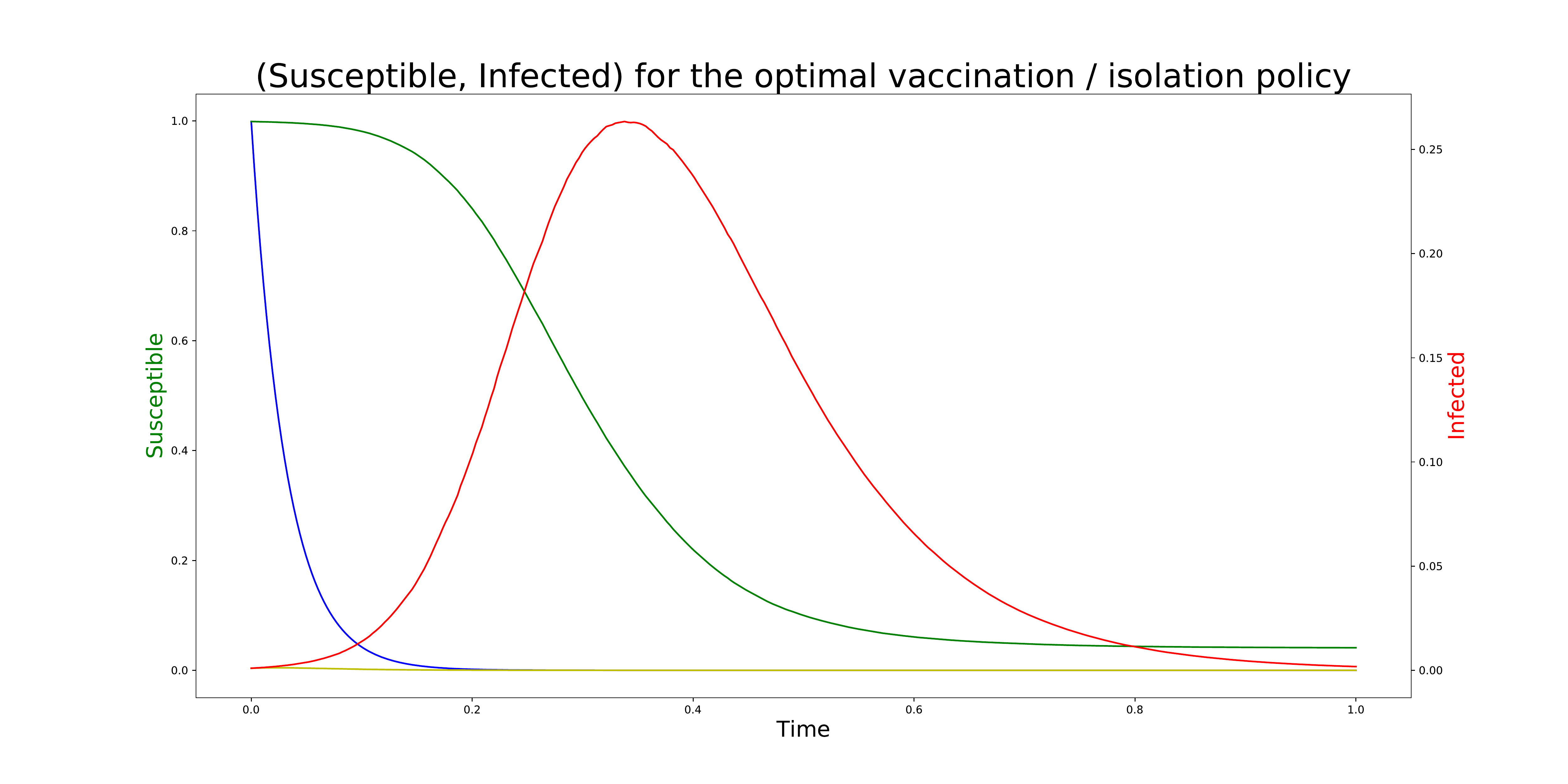}}
\caption{\label{SI_vac_iso_1}\small Raw (Susceptible, Infected) versus optimal (Susceptible, Infected) under the vaccination / isolation policy assuming high cost running cost function. The read and green lines are the average values of the raw Susceptible (primary axis) and Infected (secondary axis) fractions, respectively, while the yellow and blue lines represent the averages of the corresponding optimal values.}
\end{figure}

\begin{figure}[H]
\scalebox{1.0}[1.0]{\includegraphics[height=6.6cm]{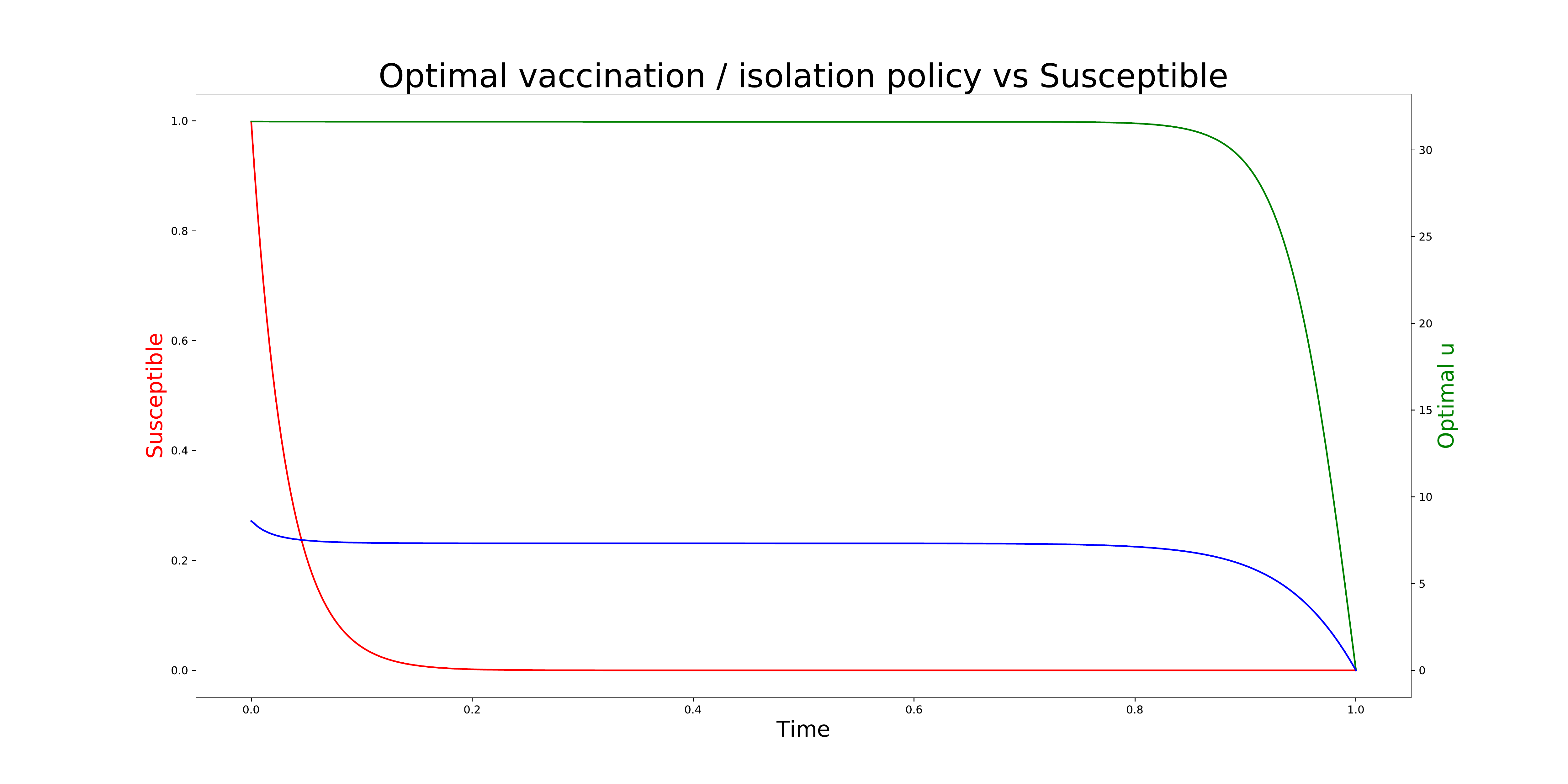}}
\caption{\label{SI_vac_iso_S_1}\small Optimal isolation (green line) and vaccination (blue line) policies (secondary axis) versus the Susceptible (red line) fraction (primary axis) for the high cost running cost function.}
\end{figure}

\begin{figure}[H]
\scalebox{1.0}[1.0]{\includegraphics[height=6.6cm]{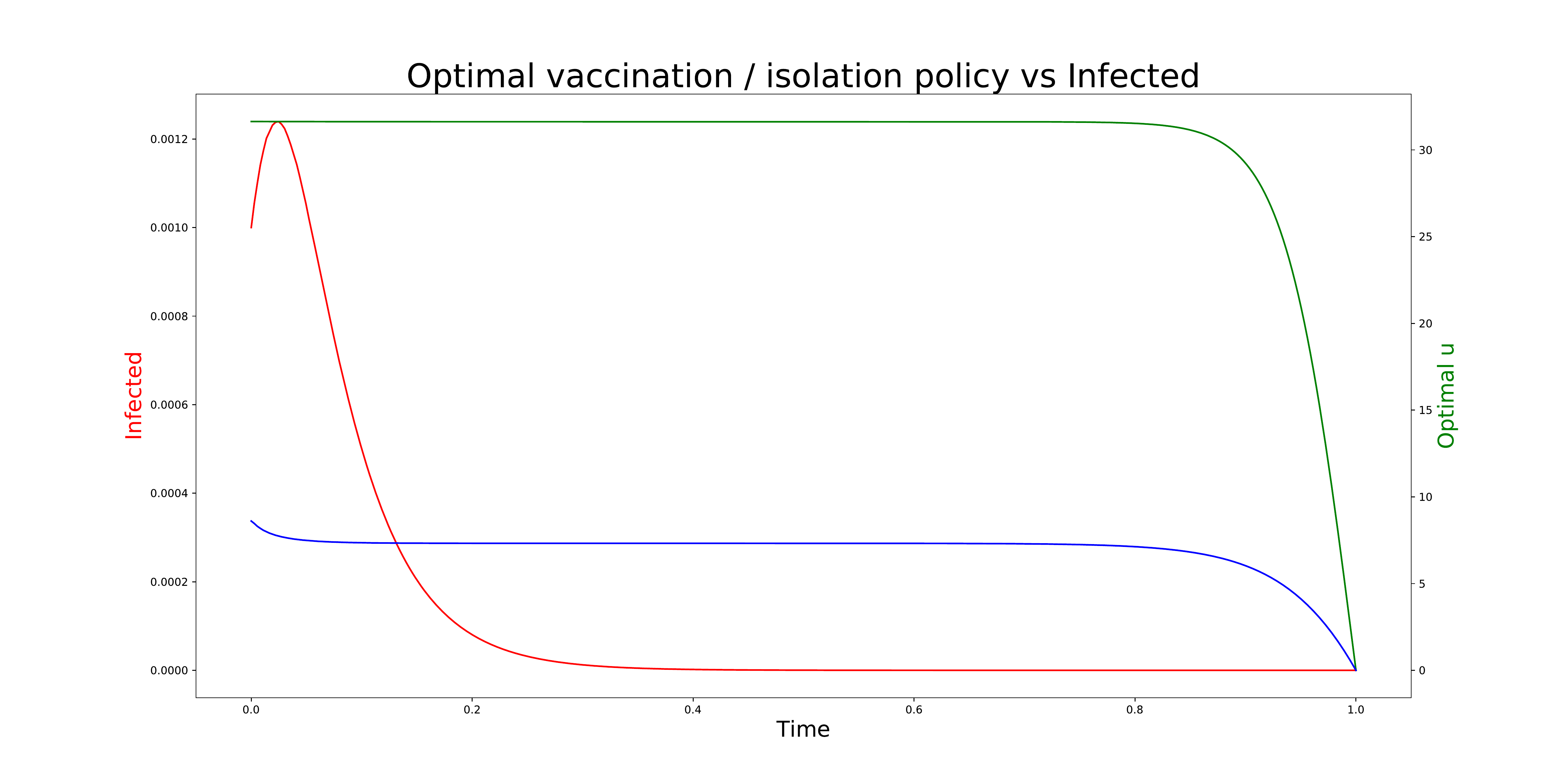}}
\caption{\label{SI_vac_iso_I_1}\small Optimal isolation (green line) and vaccination (blue line) policies (secondary axis) versus the Infected (red line) fraction (primary axis) for the high cost running cost function.}
\end{figure}

\begin{figure}[H]
\scalebox{1.0}[1.0]{\includegraphics[height=6.6cm]{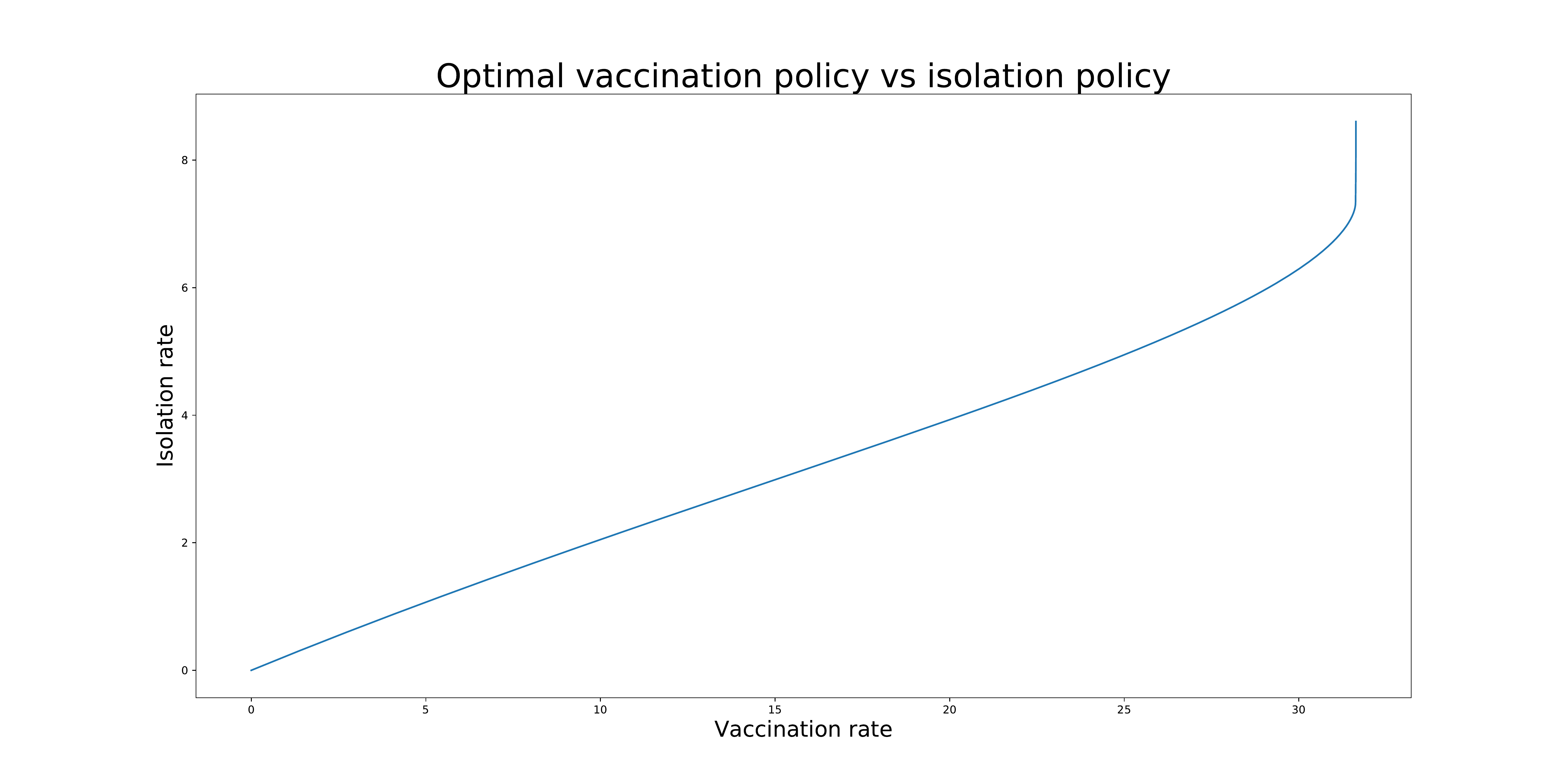}}
\caption{\label{SI_iso_vs_vac_1}\small Plot of the optimal isolation policy versus optimal vaccination policy under high cost running cost function.}
\end{figure}

Finally, we consider the case of the vaccination / isolation policy. An example of a cost running cost function is given by the following parameter values:
\begin{equation}
\begin{split}
L_1&=1.0,\\
M_1&=0.0,\\
N_1&=500.0,\\
L_2&=5.0,\\
M_2&=0.0,\\
N_2&=555.0.
\end{split}
\end{equation}
The results of Monte Carlo simulations are plotted in Figures \ref{SI_vac_iso_1} - \ref{SI_iso_vs_vac_1}.

\begin{figure}[H]
\scalebox{1.0}[1.0]{\includegraphics[height=6.6cm]{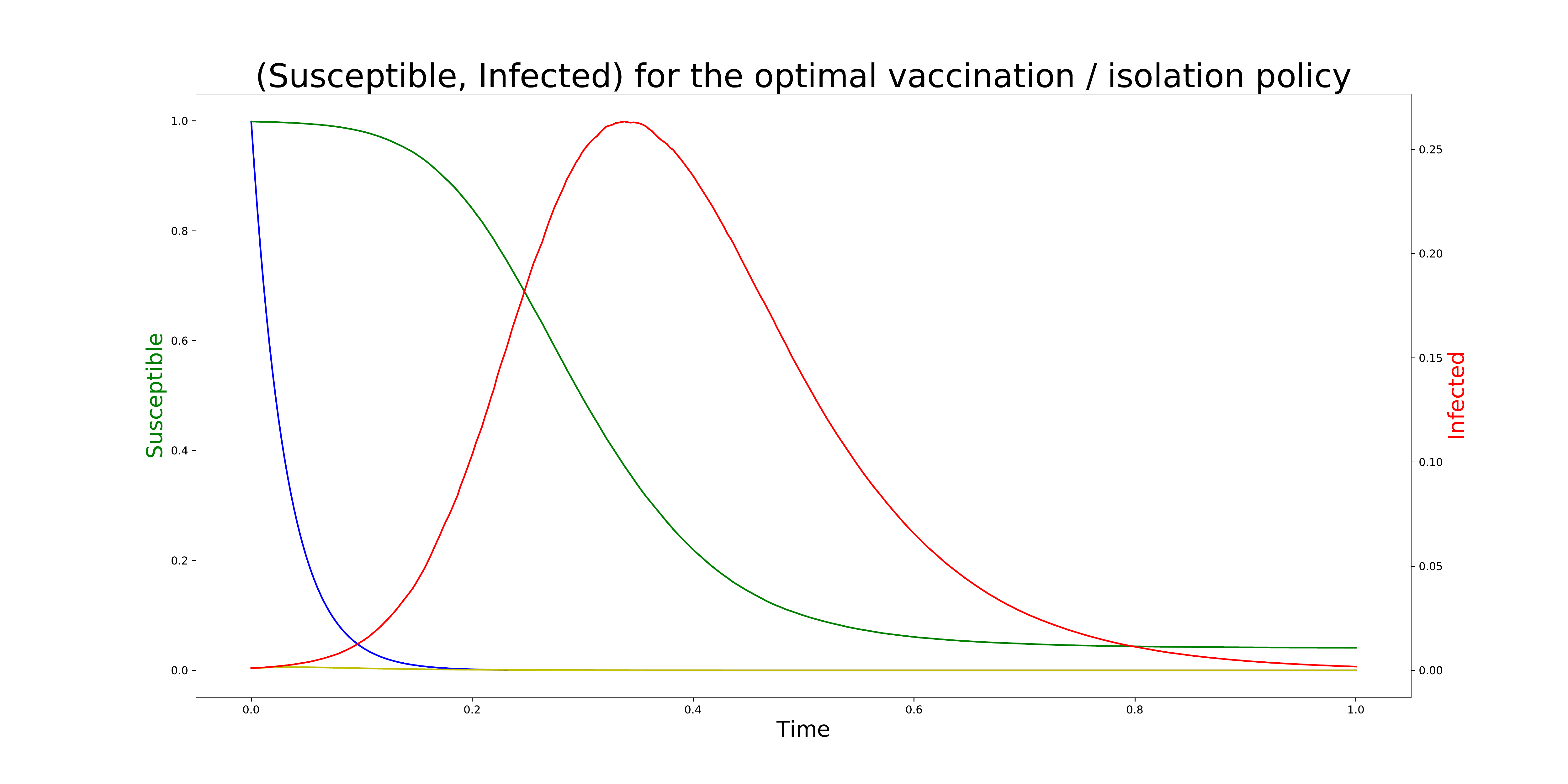}}
\caption{\label{SI_vac_iso_2}\small Raw (Susceptible, Infected) versus optimal (Susceptible, Infected) under the vaccination / isolation policy assuming low cost running cost function. The read and green lines are the average values of the raw Susceptible (primary axis) and Infected (secondary axis) fractions, respectively, while the yellow and blue lines represent the averages of the corresponding optimal values.}
\end{figure}

\begin{figure}[H]
\scalebox{1.0}[1.0]{\includegraphics[height=6.6cm]{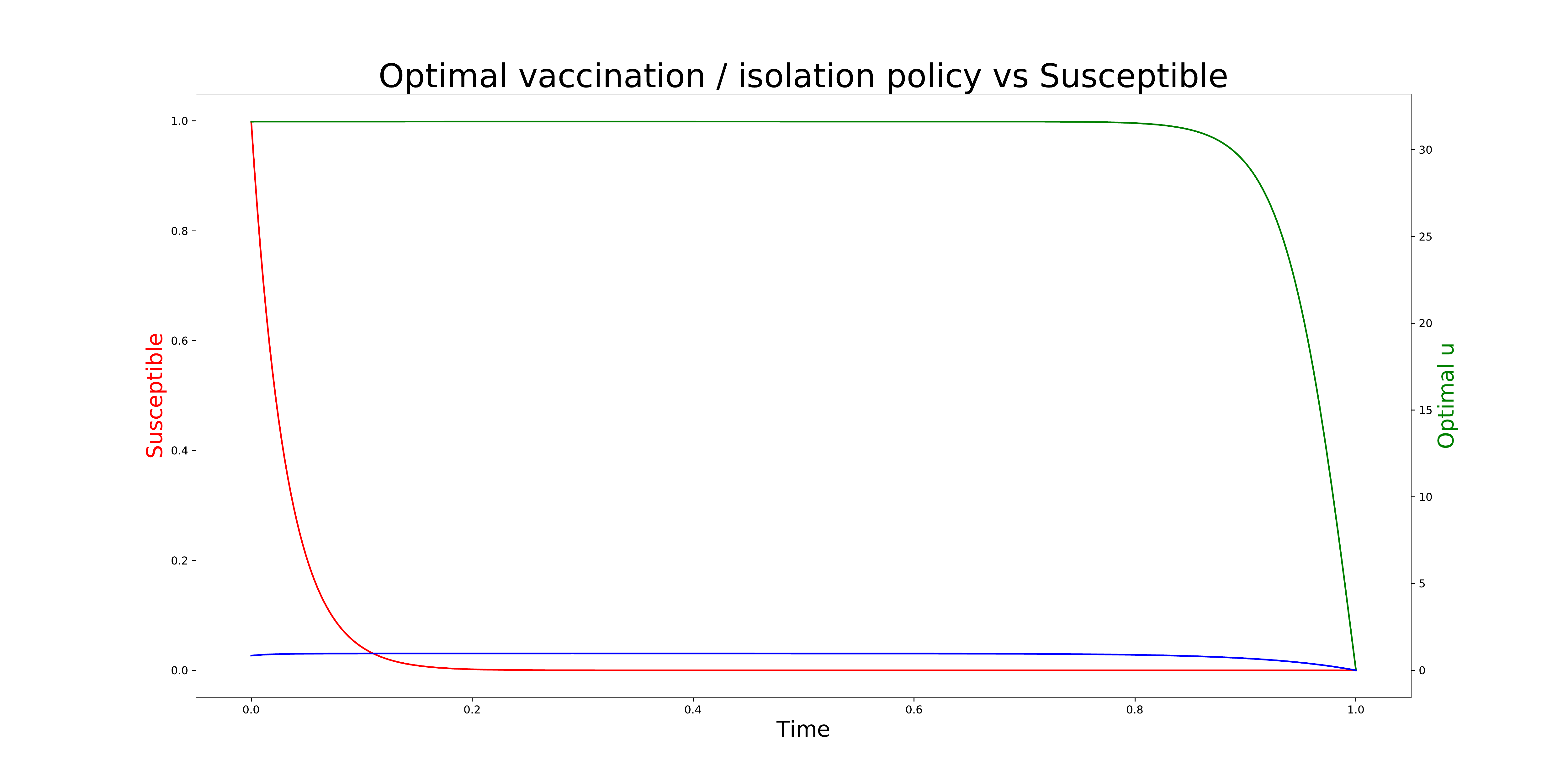}}
\caption{\label{SI_vac_iso_S_2}\small Optimal isolation (green line) and vaccination (blue line) policies (secondary axis) versus the Susceptible (red line) fraction (primary axis) for the low cost running cost function.}
\end{figure}

\begin{figure}[H]
\scalebox{1.0}[1.0]{\includegraphics[height=6.6cm]{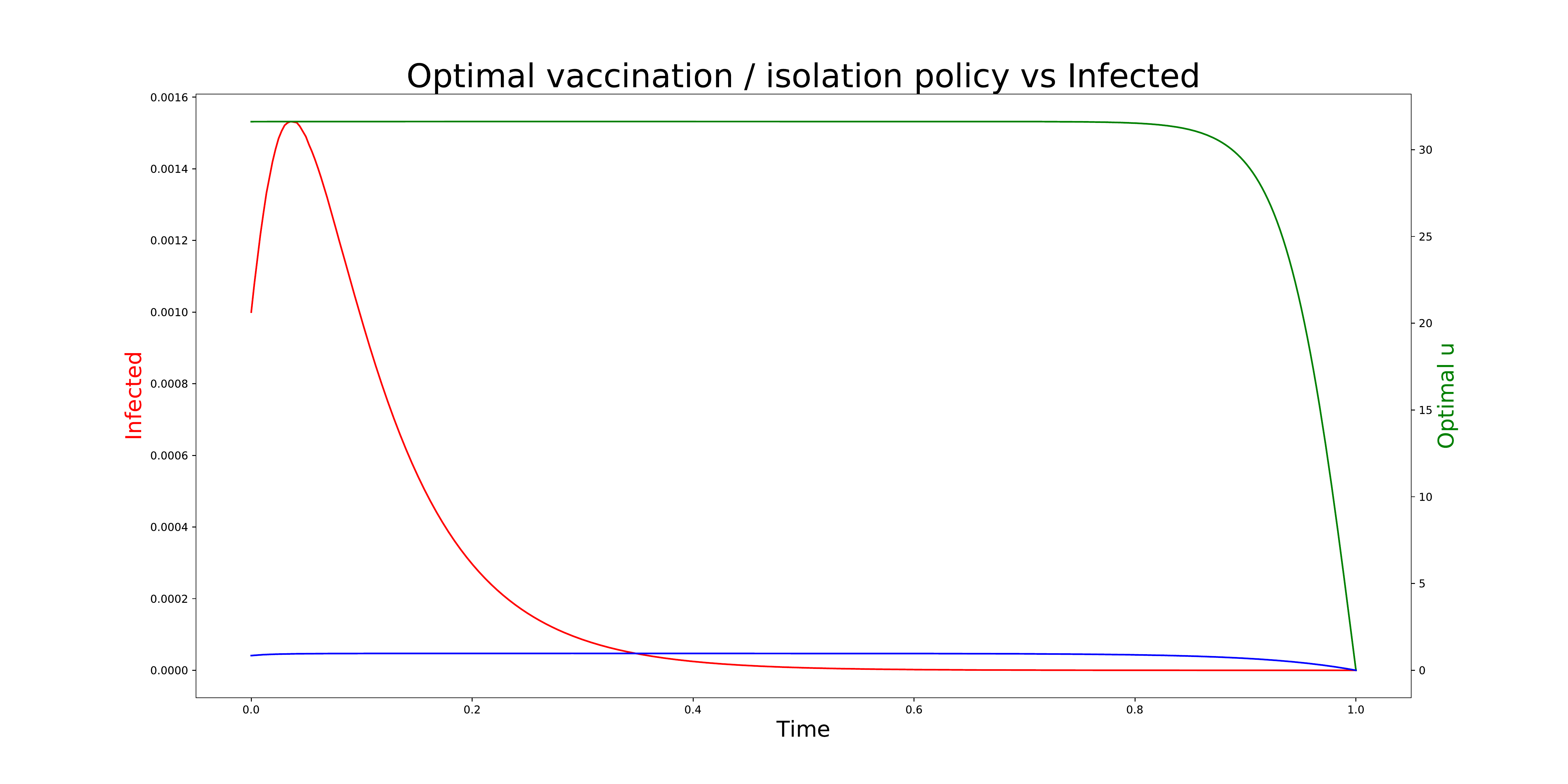}}
\caption{\label{SI_vac_iso_I_2}\small Optimal isolation (green line) and vaccination (blue line) policies (secondary axis) versus the Infected (red line) fraction (primary axis) for the low cost running cost function.}
\end{figure}

\begin{figure}[H]
\scalebox{1.0}[1.0]{\includegraphics[height=6.6cm]{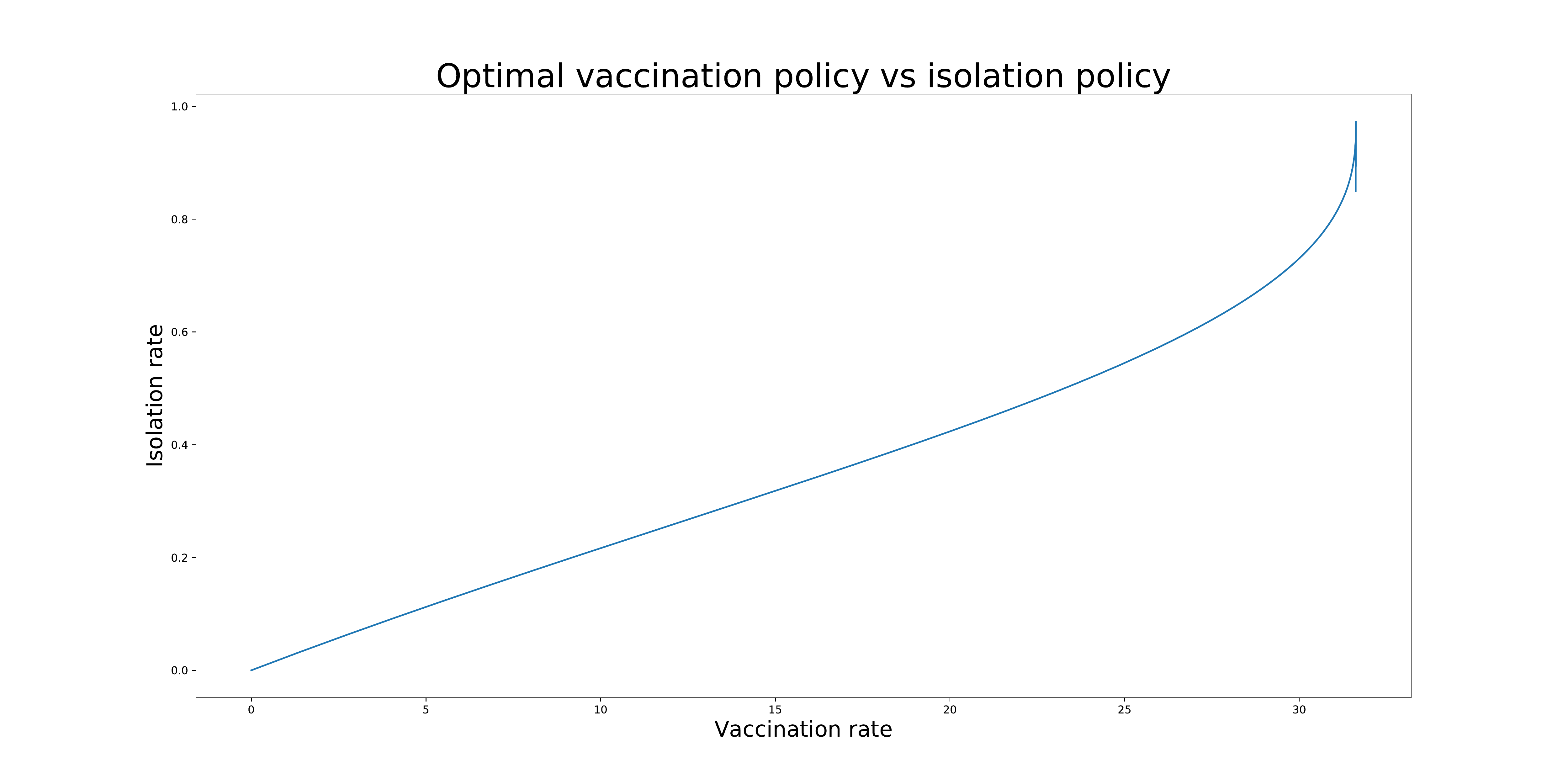}}
\caption{\label{SI_iso_vs_vac_2}\small Plot of the optimal isolation policy versus optimal vaccination policy under low cost running cost function.}
\end{figure}

A low cost running cost function is given by the following set of coefficients:
\begin{equation}
\begin{split}
L_1&=0.1,\\
M_1&=0.0,\\
N_1&=50.0,\\
L_2&=3.0,\\
M_2&=0.0,\\
N_2&=35.0.
\end{split}
\end{equation}
The results of Monte Carlo simulations are plotted in Figures \ref{SI_vac_iso_2} - \ref{SI_iso_vs_vac_2}.

\end{document}